\documentclass{amsart}
\usepackage{graphicx}
\newtheorem{thm}{Theorem}[section]

\newtheorem{lem}[thm]{Lemma}

\theoremstyle{definition}

\theoremstyle{remark}

\newtheorem{exm}[thm]{Example}
\numberwithin{equation}{section}
\newcommand{\qudit}[1]{\left\vert #1 \right\rangle}

\newcommand{\Z}{\mathbb{Z}}

\newcommand{\C}{\mathbb{C}}

\usepackage{amssymb}
\font\euler=msbm7  

\usepackage{amssymb, color}

\begin{document}

\title[]{On Binomial Summations and a Generalised Quantum SWAP Gate}

\author{Colin Wilmott and Peter Wild}
\address{School of Mathematical Sciences\\University College Dublin\\ Dublin 4, Ireland}
\address{Department of Mathematics\\ Royal Holloway\\ University of London\\ Egham\\ Surrey\\ TW20 0EX\\UK}
\thanks{\emph{Electronic address:} cmwilmott@maths.ucd.ie}
\thanks{\emph{Electronic address:} P.Wild@rhul.ac.uk}


\begin{abstract}
We give a quantum gate construction - composed entirely from
incidents of the {\small{CNOT}} gate - that generalises the qubit
{\small{SWAP}} gate to higher dimensions. This new construction is
 more regular than and is an improvement on the WilNOT quantum
gate construction.
\end{abstract}
\maketitle

\section{Introduction}

It is becoming increasingly evident that much effort continues to
be made  into finding optimal quantum circuits in the sense that
for the given  gate library there is no smaller circuit that
achieves the same task. A reason for this concerted effort is
primarily due to quantum error which is  called {decoherence}.
While a quantum computer is predicated on the undisturbed
evolution of quantum coherences,  decoherence represents a major
and unavoidable problem for the practical realisation of quantum
computers as  it describes the error state of a quantum computer
which is introduced through the interaction of system and
environment. As with classical computations, the properties of
experimentally realisable quantum gates  influence the execution
time of quantum algorithms. It is therefore advantageous to ensure
a minimal use of computational resources to limit execution time
and exposure to the quantum environment. Thus, a minimal use of
computational resources serves to cap the total decohering time
delivered by the execution of  quantum gates.

While universality of quantum gate constructions is a  key
concern, researchers now focus on searching for   universal
$n$-qubit gates that contain  fewest uses of {\small{CNOT}} gates
(Bergholm  \emph{et al.} (2004, 2005), Sedl\'ak M and Plesch
(2004), Shende \emph{et al.} (2006)). This is because the cost of
experimentally realising a {\small{CNOT}} gate exceeds the cost
associated to single qubit gates (Bergholm
 \emph{et al.} (2004)).
 Consequently, construction of quantum circuits that minimise the
 use of {\small{CNOT}} gates are important from the point of
 view of execution time of the corresponding quantum algorithm which is
  positively correlated with the decohering time of the circuit.
Among   universal designs for two-qubit quantum gates there are
those containing  six, four and three {\small{CNOT}} gates
(Bergholm
 \emph{et al.} (2004), Vatan and Williams (2004), Vidal and Dawson (2004).  However, characterising the exact
{\small{CNOT}} complexity of an arbitrary $n$-qubit operation and
constructing the corresponding efficient quantum circuit is seen
as an ambitious task, even by numerical analysis (Vidal and Dawson
(2004)), but
 researchers have done considerable work optimising their constructions (Nielsen (2005)).
In particular, Vatan and Williams (2004)  give   a quantum circuit
construction for a general two-qubit operation that requires at
most three {\small{CNOT}} gates and fifteen one-qubit gates and
show that their construction is optimal.   Crucial to in securing
this result  is the demand that quantum circuits for the two-qubit
{\small{SWAP}} gate composed using {\small{CNOT}} gates requires
at least three such gates - to show the correctness of this claim,
Vatan and Williams utilize the notion of entangling power
introduced by Zanardi \emph{et~al.} (2000). Importantly, a scheme
to realise the quantum {\small{SWAP}} gate qubits was presented by
Liang and Li where it is maintained that experimentally realising
the quantum {\small{SWAP}} gate is a necessary condition for the
networkability of quantum computation (Liang  and  Li (2005)).

Although universal two-qubit circuits with fewest uses   of
{\small{{\small{CNOT}}}} gates are known, the three-qubit Toffoli
gate is known to require at most six {\small{{\small{CNOT}}}}
gates (Nielsen and Chuang (2000)) but only five
{\small{{\small{CNOT}}}} gates have been shown to be necessary
(Vidal and Dawson (2004)); see also references therein.
Constructing a quantum circuit for a given $n$-qubit operation is
seen as an important task for the complete networkability of
quantum computation. Unfortunately, the efficiency of universal
designs that accomplish this task is often known for the worst
case of $n$-qubit operators. Although, the asymptotic number of
{\small{{\small{CNOT}}}} gates used by Barenco's decomposition to
implement any $n$-qubit operation is O($n^34^n$) (Sedl\'ak  and
Plesch (2004)), it is believed that there are interesting
operations which might require a polynomial number of
{\small{{\small{CNOT}}}} gates.

Devising quantum gate constructions that extend to $n$-qubits
remains an open problem. Furthermore,  it is believed that the
study of quantum circuit minimisation and generalisation of qubit
circuit architectures to qudit circuit architectures will require
different construction techniques than those presently known.
There is another direction in which this problem may be taken.
This is the generalisation from using qubit gates to using qudit
gates for circuit design. To this end, a geometric approach to
quantum circuit minimisation has been put forward by Nielsen
(2005) that seeks to find the length of a minimal geodesic  with
respect to a suitable Finsler metric. A striking feature of this
approach is that once an initial position and velocity of the
geodesic are determined the remainder of the geodesic can be
completely evaluated by  a second order differential equation.
This is in contrast with the usual case of classical and quantum
circuitry design where part of a circuit does not aid the complete
design.

Considering quantum network designs within the qudit setting taken
with new approaches to circuitry design ought to be merit in
itself. Therefore,  a question that remains is whether a
generalised {\small{{\small{SWAP}}}} gate for higher dimensional
quantum systems can  be completely and optimally constructed from
uses of a generalised {\small{{\small{{\small{{\small{CNOT}}}}}}}}
gate. Should this be the case then such a construction may be a
cornerstone in the networkability of quantum computers based on
qudits.

\section{Preliminaries}

Given an arbitrary finite alphabet $\Sigma$ of cardinality ${d}$,
we  process quantum information by specifying a state description
of a finite dimension quantum space, in particular, the state
description of the Hilbert space $\C^{{d}}$. While the state of an
$d$-dimensional Hilbert space can be more generally expressed as a
linear combination of  basis states $\qudit{\psi_i}$, we write
each orthonormal basis state of the ${d}$-dimensional Hilbert
space $\C^{{d}}$  to correspond with an element of $\Z_{{d}}$. In
this context the basis $\{\qudit{0}, \qudit{1}, \dots,
\qudit{{d}-1}\}$ is referred to as the \emph{computational basis}.
Therefore, a state $\qudit{\psi}$ of $\C^{{d}}$ is given by
$\qudit{\psi} = \sum_{i=0}^{{d}-1}{}\alpha_i\qudit{i}$, where
$\alpha_i \in \C$ and
$\sum_{i=0}^{\textbf{d}-1}{}{\vert\alpha_i\vert}^2 = 1$. A qu{d}it
describes a state in the Hilbert space $\C^{{d}}$, and  the state
space of an  $n$-qu{d}it state is the tensor product of the basis
states of the single system $\C^{{d}}$, written ${\mathcal{H}} =
({\C^{{d}}})^{\otimes n}$, with corresponding orthonormal basis
states given by
$\qudit{i_1}\otimes\qudit{i_2}\otimes\dots\otimes\qudit{i_n} =
\qudit{i_1i_2\dots i_n}$, where $i_j \in \Z_{{d}}$.  The general
state of a qu{d}it in the Hilbert space ${\mathcal{H}}$ is then
written
\begin{eqnarray}
\qudit{\psi} = \sum_{(i_1i_2\dots i_n) \ \in \
\textrm{\euler{Z}}_{{d}}^{n} }{}\alpha_{(i_1i_2\dots
i_n)}\qudit{i_1i_2\dots i_n},
\end{eqnarray} where $\alpha_{(i_1i_2\dots i_n)}  \in  \C$
and $\sum{}\vert\alpha_{(i_1i_2\dots i_n)}\vert^2 =1$. The qudit
representation of a quantum state provides a natural mechanism by
which quantum computations can be implemented. That such a
computation is made  possible initially lies with the notion of
state signature. In particular, the correspondence of quantum
information $\alpha_k$ with a computational qudit basis element
$\qudit{k}$ and the subsequent genesis of the quantum state
$\sum_{k=0}^{d-1}{}\alpha_k\qudit{k}$ in the Hilbert space
$\C^{d}$. Such a correspondence between information and a Hilbert
space representation is  prerequisite to quantum computation since
the  successful transmission of any information state is
predicated on encoding the basis states associated with the
quantum information elements rather than the information itself.

Let ${\mathcal{H}}_{\mathcal{{A}}}$ and
${\mathcal{H}}_{\mathcal{{B}}}$ be two ${d}$-dimensional Hilbert
spaces with bases $\qudit{i}_{\mathcal{A}}$ and
$\qudit{i}_{\mathcal{B}}, i \in \Z_{{d}}$ respectively. Let
$\qudit{\psi}_{\mathcal{A}}$ denote a pure state of the quantum
system ${\mathcal{H}}_{\mathcal{{A}}}$. Similarly, let
$\qudit{\phi}_{\mathcal{{B}}}$ denote a pure state of the quantum
system ${\mathcal{H}}_{\mathcal{{B}}}$ and  consider an arbitrary
unitary transformation $U \in \textrm{U({d}}^2)$ acting on
${\mathcal{H}_{\mathcal{{A}}}} \otimes
{\mathcal{H}_{\mathcal{{B}}}}$. Let $U_{\tiny{\textrm{CNOT}}}$
denote a controlled-NOT (CNOT) gate that has qu{d}it
$\qudit{\psi}_{\mathcal{{A}}}$ as the control qu{d}it and
$\qudit{\phi}_{\mathcal{{B}}}$ as the target qu{d}it; then
\begin{eqnarray}
U_{\tiny{\textrm{CNOT}}}\qudit{m}_{{\mathcal{{A}}}}\otimes\qudit{n}_{{\mathcal{{B}}}}
= \qudit{m}_{{\mathcal{{A}}}}\otimes\qudit{n \oplus
m}_{{\mathcal{{B}}}}, \qquad m,n\in \Z_{d}
\end{eqnarray}
 where $i\oplus j$ denotes modulo ${d}$ addition.

\section{A Generalised Quantum SWAP Gate Construction}

Let us consider a set of $d$ qudit quantum systems, the first
system ${\mathcal{A}}_0$ prepared in the state $\qudit{e_0}_0$,
the second system ${\mathcal{A}}_1$ prepared in the state
$\qudit{e_1}_1$ and so forth, with the  final system
${\mathcal{A}}_{d-1}$ prepared in the state
$\qudit{e_{d-1}}_{d-1}$. We ask  whether or not it is possible to
construct a network to implement  a  generalised {\small{SWAP}}
gate so that in the output of the network the system
${\mathcal{A}}_0$ is in the state $\qudit{e_1}_0$, the system
${\mathcal{A}}_1$ is in the state $\qudit{e_2}_1$ and so forth,
until the system ${\mathcal{A}}_{d-1}$ is in the state
$\qudit{e_0}_{d-1}$ where $e_0, \dots, e_{d-1} \in
\{0,\dots,d-1\}$. To this end, we give the following.

\subsection{The Construction}

Let $k$ and $l$ be positive integers. For non-negative integers
$j$, let us consider  the function $f(j)={j\choose k}\
\textrm{mod}\  l$ and note that $f(j)$  is a cyclic function of
$j$. We use the periodic property  of this function to study the
computational problem associated with the construction of a
generalised {\small{SWAP}} gate for quantum systems over
dimensions $d$. In particular, we will relate the binomial
summation $a_j = \sum^{j/{d}}_{i=0}{}{j-(d-1)i\choose i}$
\textrm{mod}\ $d$
 to a quantum  network for a quantum gate,
 where such a network is constructed from the periodic application of {\small{CNOT}}
 gates connecting adjacent systems of the network.

\begin{figure}
\hskip-4.5em \setlength{\unitlength}{0.08cm} \hspace*{45mm}
\put(-59.7,133.5){$\overbrace{\qquad\qquad\ \ }^{a_j \
(\textrm{mod}\ 4)}$} \put(-59.8,130.5){$\line(1,0){24.4}$}
\put(-60,120){$e_1,e_2,e_3,e_0,e_0\oplus e_1, \ \dots$}
\put(-61,126.65){$\left\uparrow\begin{matrix} \vspace*{.3mm}
\cr\end{matrix}\right.$}
\put(-42,126.65){$\left\uparrow\begin{matrix} \vspace*{.3mm}
\cr\end{matrix}\right.$}
\put(-36.5,126.65){$\left\downarrow\begin{matrix} \vspace*{.3mm}
\cr\end{matrix}\right.$} \put(-60,110){0\ \ 0\ \ 0\ \ 1\ \ 1\ \ 1\
\ 1\ \ 2\ \ 3\ \ 0\ \ 1\ \ 3\ \ 2\ \ 2\ \ 3\ \ 2\ \ 0\ \ 2\ \ 1\ \
3\ \ 3\ \ 1\ \ 2\ \ 1\ \ 0\ \ 1\ \ 3\ \ 0\ \ 0\ \ 1}
\put(-60,100){1\ \ 0\ \ 0\ \ 0\ \ 1\ \ 1\ \ 1\ \ 1\ \ 2\ \ 3\ \ 0\
\ 1\ \ 3\ \ 2\ \ 2\ \ 3\ \ 2\ \ 0\ \ 2\ \ 1\ \ 3\ \ 3\ \ 1\ \ 2\ \
1\ \ 0\ \ 1\ \ 3\ \ 0\ \ 0} \put(-60,90){0\ \ 1\ \ 0\ \ 0\ \ 0\ \
1\ \ 1\ \ 1\ \ 1\ \ 2\ \ 3\ \ 0\ \ 1\ \ 3\ \ 2\ \ 2\ \ 3\ \ 2\ \
0\ \ 2\ \ 1\ \ 3\ \ 3\ \ 1\ \ 2\ \ 1\ \ 0\ \ 1\ \ 3\ \ 0}
\put(-60,80){0\ \ 0\ \ 1\ \ 0\ \ 0\ \ 0\ \ 1\ \ 1\ \ 1\ \ 1\ \ 2\
\ 3\ \  0\ \ 1\ \ 3\ \ 2\ \ 2\ \ 3\ \ 2\ \ 0\ \ 2\ \ 1\ \ 3\ \ 3\
\ 1\ \ 2\ \ 1\ \ 0\ \ 1\ \ 3}
\begin{picture}(95,45)(55,-11)

\put(0,20){\line(1,0){127}} \put(0,30){\line(1,0){127}}
\put(0,40){\line(1,0){127}} \put(0,10){\line(1,0){127}}

\put(-7,20){$e_2$} \put(-7,30){$e_1$} \put(-7,40){$e_0$}
\put(-7,10){$e_3$}

\put(133,10){\dots}\put(133,20){\dots}\put(133,30){\dots}\put(133,40){\dots}
\put(145,20){\line(1,0){17}} \put(145,30){\line(1,0){17}}
\put(145,40){\line(1,0){17}} \put(145,10){\line(1,0){17}}

\put(45,40){\line(1,0){10}} \put(45,30){\line(1,0){10}}
\put(45,10){\line(1,0){10}}
\put(45,20){\line(1,0){10}}

\put(10,30){\circle{4}} \put(10,40){\circle*{2}}
\put(10,28){\line(0,1){12}}

\put(20,18){\line(0,1){12}} \put(20,20){\circle{4}}
\put(20,30){\circle*{2}}

\put(30,8){\line(0,1){12}} \put(30,20){\circle*{2}}
\put(30,10){\circle{4}}

\put(40,10){\line(0,1){32}} \put(40,10){\circle*{2}}
\put(40,40){\circle{4}}

\put(50,30){\circle{4}} \put(50,40){\circle*{2}}
\put(50,28){\line(0,1){12}}

\put(60,18){\line(0,1){12}} \put(60,20){\circle{4}}
\put(60,30){\circle*{2}}

\put(70,8){\line(0,1){12}} \put(70,20){\circle*{2}}
\put(70,10){\circle{4}}

\put(80,10){\line(0,1){32}} \put(80,10){\circle*{2}}
\put(80,40){\circle{4}}

\put(90,30){\circle{4}} \put(90,40){\circle*{2}}
\put(90,28){\line(0,1){12}}

\put(100,18){\line(0,1){12}} \put(100,20){\circle{4}}
\put(100,30){\circle*{2}}

\put(110,8){\line(0,1){12}} \put(110,20){\circle*{2}}
\put(110,10){\circle{4}}

\put(120,10){\line(0,1){32}} \put(120,10){\circle*{2}}
\put(120,40){\circle{4}}

\put(150,30){\circle{4}} \put(150,40){\circle*{2}}
\put(150,28){\line(0,1){12}}

\put(160,18){\line(0,1){12}} \put(160,20){\circle{4}}
\put(160,30){\circle*{2}}

\put(15.2,52){$\left\updownarrow\begin{matrix} \vspace*{5mm}
\cr\end{matrix}\right.$}

\end{picture}
\vskip0em\caption{Binomial summation  network construction over
dimension 4.}\label{4}
\end{figure}

Fig. \ref{4} illustrates such a quantum network for
$4$-dimensional quantum states. The diagram shows the sequence of
{\small{CNOT}} gates of the network. The columns of the array on
integers (modulo $4$) describe the states of the target systems
after the corresponding {\small{CNOT}} gates have been applied, so
the arrow indicates the correspondence between the system
${\mathcal{A}}_1$ being in the state $\qudit{e_0 + e_1}_1$, where
$e_0 + e_1$ is calculated modulo $4$, and the application of the
first {\small{CNOT}} gate. The sum $e_0 + e_1$ may be represented
as the dot product of the row vector $e_0, e_1, e_2, e_3$ and the
corresponding column $(1,1,0,0)^{\tiny{T}}$. Thus the array of
integers modulo $4$ has rows indexed by $\{0,1,2,3\}$ and columns
indexed by time $t=$ $-3,-2,-1,0,1,2,3,\dots$ such that column $t
= 4s+j$ with $j \in \{0,1,2,3\}$ corresponds to system
${\mathcal{A}}_j$. We denote the entries in the array by $b_{it}$,
where $i \in \{0,1,2,3\}$, and  $t = -3,-2,$ $-1,0,1,2,3,\dots$.
Putting $a_t = \sum_{i=0}^{3}{e_1}b_{it}$, $t =
-3,-2,-1,0,1,2,3,\dots$,  then, for $t = 4s+j$ with $j \in
\{0,1,2,3\}$, the state of the system ${\mathcal{A}}_j$ is
$\qudit{a_{4s+j}}_j$ - ${\mathcal{A}}_j$ has been the target of
$s+(1-\delta_{0j})$ {\small{CNOT}} gates. In general, for a
network of $d$ systems of dimension $d$, the periodic arrangement
of {\small{CNOT}} gates with control system ${\mathcal{A}}_j$ and
target system ${\mathcal{A}}_{j+1}$ for $j = 0, \dots,d-1$, where
$j+1 = 0$ for $j = d-1$, means that the sequence $(a_t)$, where
$\qudit{a_{ds+j}}_j$ is the state of system ${\mathcal{A}}_j$ at
time $t = ds+j$ with $j \in \{0,\dots,d-1\}$ and ${\mathcal{A}}_j$
has been the target of $s+ (1-\delta_{0j})$ {\small{CNOT}} gates,
satisfies the recurrence $a_{t+d} = a_{t+d-1} + a_{t}$ for all $t
\ge -d+1$, since the {\small{CNOT}} gate replaces the state
$\qudit{a_{ds+j}}_j$ of ${\mathcal{A}}_j$ with
$\qudit{a_{d(s+1)+j}}_j = \qudit{a_{ds+j}+a_{d(s+1) + j -1}}_j$.
With initial states $\qudit{a_{-d+1}}_1 =
\qudit{e_1}_1,\dots,\qudit{a_{-1}}_{d-1}$ =
$\qudit{e_{d-1}}_{d-1}$, $\qudit{a_0}_0 = \qudit{e_0}_0$ the terms
of the sequence $(a_t)$ may be written as $a_t =
\sum_{i=0}^{d-1}{e_ib_{it}}$ for sequences $(b_{it})$ which
satisfy the recurrences $b_{i(t+d)} = b_{i(t+d-1)} + b_{it}$,
$i=0,\dots,d-1$.  Indeed
\begin{eqnarray}
a_{t+d-1}+a_{t} &=& \sum_{i=0}^{d-1}{e_ib_{i(t+d-1)}} + \sum_{i=0}^{d-1}e_ib_{it}\nonumber\\
&=& \sum_{i=0}^{d-1}{e_i(b_{i(t+d-1)} + b_{it})}
\end{eqnarray}
and
\begin{eqnarray}
a_{t+d} &=& \sum_{i=0}^{d-1}{e_ib_{i(t+d)}}.
\end{eqnarray}
We also note that, for $i = 0,\dots,d-2$, the sequence $(b_{it})$
is a translate of the sequence $(b_{(i+1)t})$ by $1$ place.
Indeed these sequences are all translates of one another. We begin
by considering the solution of the recurrence relation $a_{(t+d)}
= a_{t+d-1} + a_t$. This turns out to be $a_t =
\sum_{i=0}^{t/d}{j-(d-1)i \choose i}$.

\begin{lem}\label{plm}(Rosen (2000))
${x \choose i} + {x \choose i+1} = {x+1 \choose i+1}$.
\end{lem}

\begin{lem}
Let $d$ be a positive integer. The sequence $(a_n)$ defined by
$a_n = \sum^{n/d}_{i=0}{}{n-(d-1)i\choose i}$ satisfies the
recurrence relation $a_{n+d} = a_{n+d-1} + a_n$ with initial
conditions $a_0 = \dots = a_{d-1} =1$.
\end{lem}
\emph{Proof:} Clearly the sequence $(a_n)$ as defined satisfies
$a_0=\dots=a_{d-1}=1$. Let $l$ be a non-negative integer and let
$m \in \{0,\dots,d-1\}$. Then \begin{eqnarray} a_{ld+m} =
\sum^{l}_{i=0}{}{(l-i)d+m+i\choose i},\end{eqnarray} and
\begin{eqnarray}
a_{ld+m+d-1} = a_{(l+1)d+m-1} &=&
\sum^{l+1}_{i=0}{}{(l+1-i)d+m+i-1\choose i} \nonumber\\ &=& 1 +
\sum^{l+1}_{i=1}{}{(l+1-i)d+m+i-1\choose i}\nonumber\\ &=&  1 +
\sum^{l}_{i=0}{}{(l-i)d+m+i\choose i+1}.\end{eqnarray} Hence, by
lemma \ref{plm},  we have \begin{eqnarray}a_{ld+m} + a_{ld+m+d-1}
&=& 1+ \sum^{l}_{i=0}{}{(l-i)d+m+i+1\choose i+1}\nonumber\\ &=& 1
+ \sum^{l+1}_{i=1}{}{(l+1-i)d+m+i\choose i}\nonumber\\ &=&
\sum^{l+1}_{i=0}{}{(l+1-i)d+m+i\choose i}\nonumber\\ &=&
a_{(l+1)d+m}\end{eqnarray} as required.

\noindent The solution sequence $(a_j)$ is periodic since the
recurrence relation is reversible and the sequence must repeat as
soon as $d$ consecutive terms, of which there are only finitely
many possibilities, are repeated. To determine the period of this
sequence we need some combinatorial results.

\begin{lem}\label{poi}(Rosen (2000)) $\sum_{i=0}^{k}{j+i\choose i} = {j+k+1\choose k}$ for all $j \ge 0$.
\end{lem}
\begin{lem}\label{peter3}
Let $p$ be prime and let $j \ge -1$. Then for $j = p-1\
(\textrm{mod}\ p)$, we have  ${p+j \choose p-1} = 1 \
(\textrm{mod}\ p)$ and for $j = 0,\dots,p-2 \ (\textrm{mod}\ p)$,
we have ${p+j\choose p-1} = 0  \ (\textrm{mod} \ p)$.
\end{lem}
\emph{Proof:} Let us write ${p+j\choose p-1}$ as a quotient of
factorials, and consequently, we have it that ${p+j\choose p-1} =
\frac{(p+j)(p+j-1)\dots(p)}{(j+1)!}$. Since there is a multiple of
$p$ in the numerator not cancelled by a factor in the denominator
except when $j=p-1\ (\textrm{mod}\ p)$, the result follows.

\begin{thm}(Lu and Tsai (2000)) \label{LT} Let $p$ be a prime  number and let $a$, and  $k$ be any positive integers. The integer function ${j \choose k}$ modulo $p^a$ for $j \ge k$
has  the cycle length $p^{a+e}$ where $e =
\lfloor\textrm{log}_pk\rfloor$.
\end{thm}

\noindent We now determine the period of the solution sequence $(
a_j)$ when $d$ is prime.

\begin{thm}\label{cyc} Let $d$ be a prime and consider the sequence $(a_j)$ defined by
$a_j = \sum^{j/d}_{i=0}{}{j-(d-1)i\choose i} \ \textrm{mod} \ d$.
Then $(a_j)$ has the cycle length $d^2-1$.
\end{thm}

\emph{Proof:} The sequence $(a_j)$ satisfies the recurrence
$a_{j+d} = a_{j+d-1}+a_j$ for all $j$. Since $a_0 = a_1 = \dots =
a_{d-1} = 1$, it is sufficient to show that $a_{d^2-1}=1,
a_{d^2-2}=0, \dots, a_{d^2-d}=0$. This implies that $a_{j+d^2-1} =
1$ for $j=0,\dots,d-1$, and so $a_{j+d^2-1}=a_j$ for all $j \ge
0.$ Now, for $j = 0,\dots, d-2$,
\begin{eqnarray} a_{j+d^2-d} &=&
\sum_{i=0}^{\frac{j+d^2-d}{d}}{}{j+d^2-d-(d-1)i\choose
i}\nonumber\\ &=& \sum^{d-1}_{i=0}{}{j+i\choose i}.\end{eqnarray}

\noindent By lemma \ref{poi},
 \begin{eqnarray}
\sum^{d-1}_{i=0}{}{j+i\choose i} &=& {j+d\choose d-1} \ ({\rm mod} \ d)\nonumber\\
&=& \left\{\begin{matrix} \vspace*{4mm}
\cr\end{matrix}\right.\begin{matrix}0\ \textrm{mod}\
d&\textrm{for}\ \  0\leq j<d-1\cr
    1\ \textrm{mod}\ d&\textrm{for}\ \ j=d-1.\cr\end{matrix}
\end{eqnarray}
Thus the period divides $d^2-1$. Next, we show that $P_a = d^2-1$
is the smallest cycle length. For $i=0$, ${{j-(d-1)i} \choose
i}=1$ for all $j\ge 0$. Let $i$ be such that $1 \le I <
\frac{d^2-1}{d}$. Then the sequence $(c_j)$ with $c_j={{j-(d-1)i}
\choose i}$ satisfies $c_j=0$ for $j=0,\dots,di-1$ and $c_j=1$ for
$j=di$ and is periodic with period $d$ for $j \ge di$ by Theorem
\ref{LT}. Thus, $a_{ld}=l+1$ for $l=0,\dots,d-1$. Now, any $d$
consecutive terms of $a_0,\dots,a_{d^2-1}$ includes a term
$a_{ld}$ for some $l$ and thus there    cannot be $d-1$
consecutive $0$ and one $1$ until the terms from $d^2-d$ to
$d^2-1$. This establishes the result.
\\

\noindent Since the cycle length $d^2-1$ is coprime to $d$,  note
that at the completion of a cycle the states of the system will be
cycled round. In fact, as $d^2-1=d-1= -1\  ({\rm mod} \ d)$ they
are shifted by one position and the network implements the
{\small{SWAP}} gate - system ${\mathcal{A}}_0$ will be in the
state $\qudit{e_1}_0$, the initial state of ${\mathcal{A}}_1$,
${\mathcal{A}}_1$ will be in the state $\qudit{e_2}_1$, \dots,
${\mathcal{A}}_{d-1}$ will be in the state $\qudit{e_0}_{d-1}$.
This can be seen by the following argument. Since the sequences
$(b_{it})$, $i=0,\dots,d-1$, have cycle length $d^2-1$, we have
$(b_{0t},\dots,b_{(d-1)t})$ equal to: $(0,1,0,\dots,0)$ for
$t=d^2-d $ corresponding to system ${\mathcal A}_0$;
$(0,0,1,0,\dots,0)$ for $t=d^2-d+1$ corresponding to system
${\mathcal A}_1$; $\dots$; $(0,\dots,0,1)$ for $t=d^2-2$
corresponding to system ${\mathcal A}_{d-2}$; $(1,0,\dots,0)$ for
$t=d^2-1$ corresponding to system ${\mathcal A}_{d-1}$.

\noindent Now let us consider the prime power dimension $d=p^m$
with $p$ prime and let us further  consider the  family of
binomial summations $\sum^{j/{p^m}}_{i=0}{}{j-(p^m-1)i\choose i}$.
We conjecture that the cycle length of the integer sequence
$(a_j)$ for  $a_j = \sum^{j/{p^m}}_{i=0}{}{j-(p^m-1)i\choose i}$ $
\ \textrm{mod}\ p^m$ is $p^{m-1}(p^{2m}-1)$;
 see Table \ref{cycle}. Note that if this conjecture is true then since
gcd(${p^m}, p^{m-1}(p^{2m}-1)) = p^{m-1} \ne 1,$ the network does
not produce a {\small{SWAP}} gate when $m>1$. Although at the end
of a cycle the systems are shifted round, they are shifted
$p^{m-1}$ places and the systems return to their original states
after $p$ applications of the network. Thus, the network provides
a cyclic swap on each of $p^{m-1}$ groups of $p$ systems, the
systems of a group have indices congruent modulo $p^{m-1}$. What
distinguishes this network from $p^{m-1}$ copies of the
{\small{SWAP}} network for $d = p$ is that in this larger network
the systems that are swapped do not
 directly interact through a {\small{CNOT}} gate.

\noindent We turn to the problem of determining the cycle length
of the sequence $(a_j)$ for $d=p^m,\ (m>1)$. We have it that over
such dimensions the sequence $(a_j)$ satisfies the recurrence
$a_{j+p^m} = a_{j+p^m-1}+a_j$ for all $j$. Since
$a_0=a_1=\dots=a_{p^m-1}=1$ by definition, it is sufficient  to
show that $a_{p^{m-1}(p^{2m}-1)}=1$ and $a_{p^{m-1}(p^{2m}-1)-j}$
vanishes for $j= 1,\dots,p^{m}-1$.  It then follows that
$a_{p^{m-1}(p^{2m}-1)+j}=1$ for $j=0,\dots,p^m-1$. Considering
this problem, we seek a closed form expression  for the integer
sequence $a_j = \sum_{i = 0}^{j/p^m}{{j-(p^m-1)i\choose i}}$. We
outline the approaches we have taken. Firstly, following the
approach for $d=p$, we find that
\begin{eqnarray}
a_{p^{m-1}(p^{2m}-1)-j} &=&
\sum_{i=0}^{p^{2m-1}-1}{}{p^{m-1}(p^{2m}-1)-j-(p^m-1)i\choose i}\
\textrm{mod}\ p^m\nonumber\\&=&
\sum_{i=0}^{p^{2m-1}-1}{}{(p^{m}-1)(p^{2m-1}+p^{m-1}-i)-j\choose
i}\ \textrm{mod}\ p^m\nonumber\\ &=&
\sum_{i=0}^{p^{2m-1}-1}{}{(p^{m}-1)(p^{m-1}-i)-j\choose i}\
\textrm{mod}\ p^m\nonumber\\ &=&
\sum_{i=0}^{p^{2m-1}-1}{}{(1-p^{m})i -p^{m-1}-j\choose i}\
\textrm{mod}\ p^m.
\end{eqnarray}
We ask if there exists a combinatorial approach which illustrates
that the sum $\sum_{i=0}^{p^{2m-1}-1}{}{(1-p^{m})i
-p^{m-1}-j\choose i}$ mod $p^m$ equals 1 for $j=0$ and vanishes
for $j = 1,\dots,p^m-1$. On the other hand, many combinatorial
problems have been resolved through the theory of hypergeometric
series (Petkovsek \emph{et al.}). A {hypergeometric} series
$\sum_{k\ge 0}{t_k}$ is one in which $t_0 = 1$ and the ratio of
two consecutive terms is a rational function of the summation
index. The process of identifying a given hypergeometric series by
writing the series in the standard $_{p}F_{q}\left[\cdot\right]$
may may aid a simple closed form for the series by comparing it
with the library of closed forms associated with hypergeometric
series listed in Petkovsek \emph{et al.}. Interestingly, the
sequence $(a_j)$ can be identified as a hypergeometric series of
the form
\begin{eqnarray}{}_{p^m}F_{p^m-1}\begin{bmatrix}
  \frac{-j}{p^m} & \frac{-j+1}{p^m}, & \dots,  \left(\frac{-j+p^m-1}{p^m}\right) & & \\
&  & & ;-\frac{(p^{m})^{p^m}}{(p^{{m-1}})^{p^{m-1}}}&\\
\frac{-j}{p^m-1},&\frac{-j+1}{p^m-1},& \dots,  \frac{-j+p^m-2}{p^m-1} & & \\
\end{bmatrix}.\end{eqnarray}
Unfortunately, the library of closed forms  listed in Petkovsek
\emph{et al.}  does not provide a closed form for the sum
$\sum^{j/{p^m}}_{i=0}{}{j-(p^m-1)i\choose i}$. Instead we consider
the following approach for evaluating the closed form of the
binomial summation  $ a_j =
\sum^{j/{p^m}}_{i=0}{}{j-(p^m-1)i\choose i}$ mod $p^m$.

\noindent Let $A(z)$ be the power series $\sum_{j\geq0}^{}{a_j
z^j}$ and denote by $[z^j]A(z)$ the coefficient of $z^j$ in
$A(z)$; thus $[z^j]A(z) = a_j$. Now we determine the
\emph{generating function} for $A(z)$. A generating function is a
clothesline on which we hang up a sequence of numbers for display
(Wilf). In particular, the $j$th term of the sequence $(a_j)$ is
the coefficient of $z^{j}$ in the expansion of its generating
function as a power series. To find the generating function
associated the sequence $(a_j)$, we first note that the sequence
$(a_j)$ satisfies the recurrence relation
\begin{eqnarray}a_j = \left\{\begin{matrix}
0& \hskip-5em\textrm{if} \ \ \ j<0\cr 1& \textrm{if}\ \ \  j =
0,\dots,p^m-1\cr a_{j-1} + a_{j-p^m}&
\hskip-5em\textrm{otherwise.}\cr
\end{matrix}\right.\end{eqnarray}
This recurrence relation can be expressed as the single equation
\begin{eqnarray}\label{tyu}
a_j = a_{j-1} + a_{j-p^m} + [j=0]
\end{eqnarray}
where $[j=0]$ adds 1 when $j=0$. To demonstrate the generating
function for  $A(z)$, we multiply both sides of  (\ref{tyu}) by
$z^j$ and sum over $j$. Thus, we find
\begin{eqnarray}
\sum_{j=0}^\infty{a_jz^j} &=& \sum_{j=0}^\infty{a_{j-1}z^j}+\sum_{j=0}^\infty{a_{j-p^m}z^j}+\sum_{j=0}^\infty{[j=0]z^j}\nonumber\\
&=& \sum_{j=0}^\infty{a_{j}z^{j+1}}+\sum_{j=0}^\infty{a_{j}z^{j+p^m}}+ 1 \nonumber\\
&=& z\sum_{j=0}^\infty{a_jz^j}+z^{p^m}\sum_{j=0}^\infty{a_jz^j}+1.
\end{eqnarray}
Consequently, the generating function for $A(z)$ is given by
$1/(1-z-z^{p^m})$. We can restate our problem to find a closed
form for $A(z)$, and thus evaluate $\small{[z^j]}A(z)$.  We use
the following result from Graham \emph{et al.} (1994).

\begin{lem}(Graham \emph{et al.} (1994)) \label{IEEE} $\frac{1}{{(1-\alpha z)}^{i+1}} = \sum_{n=0}^{\infty}{{i+n \choose i}\alpha^n z^n}.$
\end{lem}

\noindent We seek ${[z^j]}A(z) = {[z^j]}{1}/{(1-z-z^{p^m})}$.
Consider the series, \begin{eqnarray}{1}/{{(1-\alpha z)}^{i+1}} =
\sum_{j=0}^{\infty}{{i+j \choose i}\alpha^j z^j},\end{eqnarray}
 and further consider   a finite sum of such series
\begin{eqnarray}S(z) = \frac{\beta_1}{{(1-\alpha_1
z)}^{i_1+1}}+\dots+\frac{\beta_N}{{(1-\alpha_N z)}^{i_N+1}}.
\end{eqnarray} Then $[z^j]S(z)$ is the finite sum of coefficients
given by
\begin{eqnarray}\label{lj}
[z^j]S(z) = \beta_1{i_1+j \choose
i_1}\alpha_1^j+\dots+\beta_N{i_N+j \choose i_N}\alpha_N^j.
\end{eqnarray}
Let $A(z) = \frac{1}{B(z)}$ where $B(z) = 1-z-z^{p^m}$. We now
show that $B(z)$ has distinct roots. Let us suppose that $B(z)$
has a set of repeated roots. Then, we have it that $B(z)$ and
$B'(z)$ share a set of common roots. Since  $B'(z) =
-1-p^m(z)^{p^m-1}$, it follows that a repeated root $\alpha$
satisfies $B'(\alpha) =  -1-p^m(\alpha)^{p^m-1} = 0$, and
consequently,     $\alpha^{{p^m}-1} = -1/p^m$. Whence,
$\alpha^{{p^m}} = -\alpha/p^m$. Furthermore, as $B(\alpha)$
vanishes, we have it that $1-\alpha+\alpha/p^m = 0$, or
equivalently, $1-\alpha(1-1/p^m) = 0$
from which we deduce  $\alpha = p^m/(p^m-1)$ to be the only candidate for a repeating root. Therefore, $\alpha = p^m/(p^m-1)$ should  satisfy the equation $\alpha^{{p^m}-1} = -1/p^m$,  as a consequence of being  a root of $B'(z)$.  That is $\frac{(p^m)^{p^m-1}}{({p^m}-1)^{p^m-1}} = \frac{-1}{p^m}$. 
Now $(p^m)^{p^m} \equiv 0\ (\textrm{mod}\ p)$ while $
-({p^m}-1)^{p^m-1} \equiv -1\ (\textrm{mod}\ p)$ implies that
$\alpha = p^m/(p^m-1)$ is not a root of $B'(z)$. Therefore, $B(z)$
has  distinct roots.

\noindent Now writing $B(z)$ in the form
 $(z-b_1)\dots(z-b_{p^m})$, and taking reciprocals $\alpha_k$ of
$b_k$, $k = 1,\dots,p^m$,  we establish a correspondence with the
polynomial $(1-\alpha_1z)\dots(1-\alpha_{p^m}z)$. Thus, $A(z) =
1/((1-\alpha_1z)\dots(1-\alpha_{p^m}z))$ may take the form
$\beta_1/(1-\alpha_1z)+\dots+\beta_{p^m}/(1-\alpha_{p^m}z)$, for
some $\beta_l, \ l = 1,\dots, p^m.$ Note $1/(1-\alpha z)$ is a
special case of Lemma \ref{IEEE} with $i=0$.

\begin{thm} We claim that
\begin{eqnarray}\label{claim1}
[z^j]A(z) &=& {\sum_{l=1}^{p^m}{}\beta_l}{\alpha_l^j},
\end{eqnarray} where $\beta_l = -\alpha_l/B'(1/\alpha_l)$.
\end{thm}
\emph{Proof:}
\begin{eqnarray}\lim_{z\rightarrow1/\alpha_l}(z-1/\alpha_l) A(z) =
\lim_{z\rightarrow1/\alpha_l}(z-1/\alpha_l) \textrm{S}(z),
\end{eqnarray} where $\textrm{S}(z)$ is the special case of
equation \ref{lj} with $i_l = 0$. Now, it  follows that
\begin{eqnarray}\label{aa}
\lim_{z\rightarrow1/\alpha_l}(z-1/\alpha_l) A(z) &=& \lim_{z\rightarrow1/\alpha_l}(z-1/\alpha_l) \frac{1}{B(z)} \nonumber\\
&=& \lim_{z\rightarrow1/\alpha_l}\frac{z-1/\alpha_l}{B(z) - B(1/\alpha_l)}\nonumber\\
&=& \frac{1}{B'(1/\alpha_l)}
\end{eqnarray}
and,
\begin{eqnarray}\label{bb}
\lim_{z\rightarrow1/\alpha_l}{(z-1/\alpha_l)}\textrm{S}(z) &=& \lim_{z\rightarrow1/\alpha_l}{(z-1/\alpha_l)}\sum_{k=1}^{p^m}{}\frac{\beta_k}{{(1-\alpha_k z)}}  \nonumber\\
&=& \lim_{z\rightarrow1/\alpha_l}\frac{\beta_l(z-1/\alpha_l)}{-\alpha_l(z-1/\alpha_l)}\nonumber\\
&=&  \frac{\beta_l}{-\alpha_l}.
\end{eqnarray}
Consequently, we deduce  $\beta_l = \frac{-\alpha_l}{B'(1/\alpha_l)}$, for $l = 1,\dots,p^m$, since \begin{eqnarray}\lim_{z\rightarrow1/\alpha_l}{(z-1/\alpha_l)}\frac{\beta_k}{{(1-\alpha_k z)}}\end{eqnarray} vanishes for $k \ne l$ and the result follows. 

\noindent We now test our closed form, equation (\ref{claim1}),
against the sequence of integers arising for instances of the
binomial summation $a_j =
\sum^{j/{p^m}}_{i=0}{}{j-({p^m}-1)i\choose i}$ for cases $p^m = 4$
and $p^m = 8$.
\begin{exm}
\underline{Case $p^m = 4.$} The reciprocals of the roots of
$1-z-z^4$ are given as \begin{eqnarray}\alpha_1 &=&
-.8191725134,\nonumber\\ \alpha_2 &=& .219447421-.9144736630
\iota,\nonumber\\ \alpha_3 &=& .219447421+.9144736630
\iota,\nonumber\\ \alpha_4 &=& 1.380277569.\nonumber\end{eqnarray}
Since $\beta_l = \frac{-\alpha_l}{B'(1/\alpha_l)}$, we
have\begin{eqnarray}\beta_1 &=&  .1305102698,\nonumber\\ \beta_2
&=& .1610008758+.1534011260 \iota,\nonumber\\ \beta_3 &=&
.1610008758-.1534011260\iota,\nonumber\\ \beta_4 &=&
.5474879784.\nonumber\end{eqnarray} Following result
(\ref{claim1}), a closed form for the binomial coefficients $a_j$
is then given by \begin{eqnarray} a_j &=&
(.1305102698)(-.8191725134)^j\nonumber\\&&+\ (.1610+.1534
\iota)(.2194-.9144 \iota)^j\nonumber\\&&+ \
(.1610-.1534\iota)(.2194+.9144 \iota)^j\nonumber\\&&+ \
(.5474879784)(1.380277569)^j. \end{eqnarray}
\underline{Maple Input:} for j from 0 to 25 do; $a_j$;  end;\nonumber\\
\underline{Maple Output:}
1,1,1,1,2,3,4,5,7,10,14,19,26,36,50,69,95,\nonumber\\
131,181,250,345,476,657,907,1252,1728.
\end{exm}
\begin{exm}
\underline{Case $p^m = 8.$} The reciprocals of the roots of $1-z-z^8$ are given as \begin{eqnarray}\alpha_1 &=& -.9115923535,\nonumber\\ \alpha_2 &=& -.6157823065-.6871957511 \iota,\nonumber\\ \alpha_3 &=& -.6157823065+.6871957511 \iota,\nonumber\\ \alpha_4 &=& .1033089835-.9564836042 \iota \nonumber\\\alpha_5 &=& .1033089835+.9564836042 \iota,\nonumber\\
\alpha_6 &=& .8522421840-.6352622030 \iota,\nonumber\\
\alpha_7 &=& .8522421840+.6352622030 \iota,\nonumber\\
\alpha_8 &=& 1.232054631.
\end{eqnarray}
The set $\beta_l, \ l = 1,\dots,8$, is\begin{eqnarray}\beta_1 &=&  .06378010282, \nonumber\\ \beta_2 &=& .06449005934+.02789285455 \iota,\nonumber\\ \beta_3 &=& .06449005934-.02789285455 \iota,\nonumber\\ \beta_4 &=& .06911712233+.06926484155 \iota,\nonumber\\
\beta_5&=& .06911712233-.06926484155 \iota,\nonumber\\
\beta_6 &=& .1188399306+.1719523210 \iota,\nonumber\\
\beta_7 &=& .1188399306-.1719523210 \iota,\nonumber\\
\beta_8 &=& .4313256714,\nonumber
\end{eqnarray}
Again a closed form for the binomial coefficients $a_j$ is given
by \begin{eqnarray} a_j &=&
(.06378010282)(-.9115923535)^j\nonumber\\&&+\ (.0644+.0278
\iota)(-.61578230-.68719575 \iota)^j\nonumber\\&&+ \ (.0644-.0278
\iota)(-.61578230+.68719575 \iota)^j\nonumber\\&&+ \ (.0691+.0692
\iota)(.10330898-.95648360 \iota)^j \nonumber\\&&+ \ (.0691-.0692
\iota)(.10330898+.95648360 \iota)^j \nonumber\\&&+ \ (.1188+.1719
\iota)(.85224218-.63526220 \iota)^j \nonumber\\&&+ \ (.1188-.1719
\iota)(.85224218+.63526220 \iota)^j \nonumber\\&&+ \
(.4313256714)(1.232054631)^j. \end{eqnarray}
\underline{Maple Input:} for j from 0 to 25 do; $a_j$;  end;\nonumber\\
\underline{Maple
Output:}1,1,1,1,1,1,1,1,2,3,4,5,6,7,8,9,11,14,18,23,29,\nonumber\\36,44,53,64,78.
\end{exm}

\noindent Having found a closed form for $A(z) =
\sum_{j\geq0}{a_jz^j}$, we now outline the classical means by
which the cycle length of $(a_j)\  \textrm{mod}\ p^m$ may be
determined. Fix a  particular value of $m>1$ and having obtained
the closed form for the coefficients of the corresponding
polynomial $A(z)$,
 evaluate the closed form of $a_j$ for $j = 0,\dots, p^{m-1}(p^{2m}-2)$. To
 show that $p^{m-1}(p^{2m}-1)$ is the cycle length of $(a_j)$ mod $p^m$,
 show that firstly $a_{p^{m-1}(p^{2m}-1)}=1$ and $a_{p^{m-1}(p^{2m}-1)-j}$ vanishes
 for $j= 1,\dots,p^{m}-1$, and lastly,  show the subsequence of $p^m$ opening 1s does not appear
 until $j$ is $p^{m-1}(p^{2m}-1)$.
Table \ref{cycle} gives the cycle length of the sequence $(a_j)$
mod $p^m$ for small values of $p^m$. Cycle lengths for all prime
powers up to 3125 have been confirmed and they all agree with the
conjecture.

We now consider the sequence $(a_j)$ mod $d$ where  $a_j =
\sum^{j/{d}}_{i=0}{}{j-(d-1)i\choose i}$ for composite $d$ with
prime factorization $d = p_1^{m_1}\dots p_{r}^{m_r}$. We calculate
the cycle length of $(a_j)$ modulo $d$ by calculating its cycle
length modulo $p_t^{m_t}$ for each $t = 1,\dots, r$. The cycle
length of $\sum^{j/{d}}_{i=0}{}{j-(d-1)i\choose i}\  \textrm{mod}\
{p_t^{m_t}}$, for $t = 1,\dots,r$, is given by the $j$ for which
$\sum^{j/d}_{i=1}{}{j-(d-1)i\choose i} \ \textrm{mod}\ p_t^{m_t}$
equals 1 and  which has a  preceding sequence
$(a_{j-d+1},\dots,a_{j-1})$ $= (0,\dots,0)\ \textrm{mod}\
{p_t^{m_t}}$. Given that the mapping $\lambda_{d,(p_1^{m_1}\dots
p_{r}^{m_r})}: \Z_{d}\mapsto
\Z_{p_1^{m_1}}\times\dots\times\Z_{p_{r}^{m_r}}$ is well-defined
then the cycle length
$\left\vert\sum^{j/{d}}_{i=0}{}{j-(d-1)i\choose i} \ \textrm{mod}\
d\right\vert$  of  the recurrence relation of $(a_j)$ mod $d$ is
given by the $\textrm{LCM}\left\{
\left\vert\sum^{j/{d}}_{i=0}{}{j-(d-1)i\choose i}\  \textrm{mod}\
p_t^{m_t}\right\vert \right\}^{r}_{t=1}$. Table \ref{cycle} gives
some initial values for the cycle length of
$\sum^{j/{d}}_{i=0}{}{j-(d-1)i\choose i} \ \textrm{mod}\ d$. We
state this result as a theorem.

\begin{thm}\label{composition}
Let $\sum^{j/{d}}_{i=0}{}{j-(d-1)i\choose i} \textrm{mod}\ d$ be
an integer sequence and consider the decomposition
$\left\{\sum^{j/{d}}_{i=0}{}{j-(d-1)i\choose i} \textrm{mod}\
{p_t^{m_t}}\right\}_{t=1}^{r}$   of
$\sum^{j/{d}}_{i=0}{}{j-(d-1)i\choose i} \textrm{mod}\ d$ into a
direct product of disjoint cycles. Let
$\left\vert\sum^{j/{d}}_{i=0}{}{j-(d-1)i\choose i} \textrm{mod}\
d\right\vert$ be the cycle length of
$\sum^{j/{d}}_{i=0}{}{j-(d-1)i\choose i} \textrm{mod}\ d$ be and
let $\left\{\left\vert\sum^{j/{d}}_{i=0}{}{j-(d-1)i\choose i}
\textrm{mod}\ {p_t^{m_t}}\right\vert\right\}_{t=1}^{r}$ be the
cycle lengths of $\left\{\sum^{j/{d}}_{i=0}{}{j-(d-1)i\choose i}
\textrm{mod}\ {p_t^{m_t}}\right\}_{t=1}^{r}$, respectively. Then,
the cycle length of $\sum^{j/{d}}_{i=0}{}{j-(d-1)i\choose i}
\textrm{mod}\ d$ is
$\textrm{LCM}\left\{\left\vert\sum^{j/{d}}_{i=0}{}{j-(d-1)i\choose
i} \textrm{mod}\ {p_t^{m_t}}\right\vert\right\}_{t=1}^{r}$.
\end{thm}

\begin{table}\vskip1em
\hskip-4.5em \setlength{\unitlength}{0.08cm} \hspace*{65mm}
\vskip-5em\begin{picture}(100,140)(-5,50)
\put(6,60){\line(0,1){98}} \put(80,60){\line(0,1){98}}
\put(20,60){\line(0,1){98}} \put(6,158){\line(1,0){74}}
\put(6,144){\line(1,0){74}}

\put(6,60){\line(1,0){74}}

\put(12,150){$d$}
\put(24,150){$\left\vert\sum^{j/{d}}_{i=0}{}{j-(d-1)i\choose i}
\textrm{mod}\ d\right\vert$} \put(12,135){$2$} \put(12,125){$3$}
\put(12,115){$4$} \put(12,105){$5$}\put(12,95){$6$}
\put(12,85){$7$}\put(12,75){$8$} \put(12,65){$9$}

\put(40,135){$3 = 2^2-1$} \put(40,125){$8 = 3^2-1$}
\put(38,115){$30 = 2(2^4-1)$} \put(38,105){$24 =
5^2-1$}\put(26,95){$\textrm{LCM}(63,728) = 6552$} \put(39,85){$48
= 7^2-1$}\put(37,75){$252 = 4(2^6-1)$} \put(37,65){$240 =
3(3^4-1)$}

\end{picture}
\caption{Cycle length of $\sum^{j/{d}}_{i=0}{}{j-(d-1)i\choose i}
\textrm{mod}\ d.$}\label{cycle}
\end{table}

The results provided in Table \ref{cycle} are the cycle lengths of
$(a_j)$ mod $d$ for some small values of $d$. The results are
two-fold; firstly, the results  give the minimum number of
{\small{CNOT}} gates required to effectuate a quantum gate
according the periodicity of the binomial summation construction.
Secondly, use of the binomial summation construction illustrates
for which dimensions the  generalised {\small{SWAP}} gate can be
realised. In particular, we have a {\small{SWAP}} gate in those
dimensions $d$ for which the cycle length of the function $(a_j)$
mod $d$ is equivalent to $-1$ mod $d$. For dimensions $d$ where
our  {\small{SWAP}} gate is not possible then the cycle length of
$(a_j)$ mod $d$ describes a permutation, other than that
permutation sought, of the $d$ input qudit states. For example, in
dimension 4 the cycle length of $(a_j)$ mod 4 is equivalent to 2
mod 4. Therefore, we have it that the quantum gate associated with
this particular instance of the binomial summation construction
evolves four 4-dimensional quantum states such that the state
$\qudit{e_0}_0$ of the first quantum system ${\mathcal{A}}_0$ is
transposed with the state $\qudit{e_2}_2$ of the third quantum
system ${\mathcal{A}}_2$ so that the quantum system
${\mathcal{A}}_0$ is in the state $\qudit{e_2}_0$ and the quantum
system ${\mathcal{A}}_2$ is in the state $\qudit{e_0}_2$.
Correspondingly, the state $\qudit{e_1}_1$ of the second quantum
system ${\mathcal{A}}_1$ is transposed with the state
$\qudit{e_3}_3$ of the fourth quantum system ${\mathcal{A}}_3$ so
that the quantum system ${\mathcal{A}}_1$ is in the state
$\qudit{e_3}_1$ and the quantum system ${\mathcal{A}}_3$ is in the
state $\qudit{e_1}_3$.  Moreover, the gate network for an
implementation of a  generalised $\textrm{SWAP}^{(l)}$ gate, for
$l = 1, \dots, d-1$, is an instance of $l$ applications of this
design - such a network cycles the initial state description
through $l$ systems.

\noindent Furthermore, the cycle length of the function $a_j =
\sum^{j/{{6}}}_{i=0}{}{j-5.i\choose i}\ \textrm{mod}\ {6}$ over
dimension 6 is the least common multiple of the cycle lengths
associated with the functions $\sum^{j/{{6}}}_{i=0}{}{j-5.i\choose
i} \textrm{mod}\ {2}$ and $\sum^{j/{{6}}}_{i=0}{}{j-5.i\choose i}
\textrm{mod}\ {3}$. The cycle lengths of
$\sum^{j/{{6}}}_{i=0}{}{j-5.i\choose i} \textrm{mod}\ {2}$ and
$\sum^{j/{{6}}}_{i=0}{}{j-5.i\choose i} \textrm{mod}\ {3}$ are 63
and 728, respectively.  By Theorem \ref{composition}, the cycle
length of $(a_j)$ mod 6 is 6552; the least common multiple of 63
and 728. However, since 6552 is equivalent to 0 mod 6, we have it
that the {\small{CNOT}} architecture associated with the binomial
summation function $\sum^{j/{{6}}}_{i=0}{}{j-5.i\choose i}
\textrm{mod}\
{6}$ acts trivially on the input state  of  six 6-dimensional quantum states. %

\section{Qutrit SWAP}
We improve the WilNOT gate construction (Wilmott and Wild (2008c))
by considering a construction based on binomial summations that
yields a quantum gate which is also composed entirely from uses of
the {\small{CNOT}} gate.
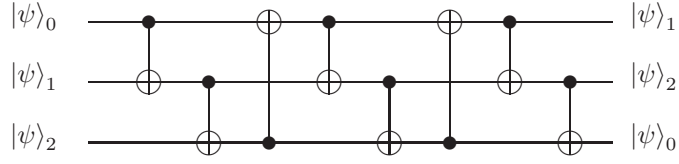
\begin{figure}\vskip1em
\hskip-4.5em \setlength{\unitlength}{0.08cm} \hspace*{25mm}
\begin{picture}(65,25)(17,20)

\put(0,20){\line(1,0){87}} \put(0,30){\line(1,0){87}}
\put(0,40){\line(1,0){87}}

\put(-13,20){$\qudit{\psi}_2$} \put(-13,30){$\qudit{\psi}_1$}
\put(-13,40){$\qudit{\psi}_0$}

\put(90,20){$\qudit{\psi}_0$} \put(90,30){$\qudit{\psi}_2$}
\put(90,40){$\qudit{\psi}_1$}

\put(45,40){\line(1,0){10}} \put(45,30){\line(1,0){10}}
\put(45,20){\line(1,0){10}}

\put(10,30){\circle{4}} \put(10,40){\circle*{2}}
\put(10,28){\line(0,1){12}}

\put(20,18){\line(0,1){12}} \put(20,20){\circle{4}}
\put(20,30){\circle*{2}}

\put(30,20){\line(0,1){22}} \put(30,20){\circle*{2}}

\put(30,40){\circle{4}}

\put(40,30){\circle{4}} \put(40,40){\circle*{2}}
\put(40,28){\line(0,1){12}}

\put(50,18){\line(0,1){12}} \put(50,20){\circle{4}}
\put(50,30){\circle*{2}}

\put(60,20){\line(0,1){22}} \put(60,20){\circle*{2}}

\put(60,40){\circle{4}}

\put(70,30){\circle{4}} \put(70,40){\circle*{2}}
\put(70,28){\line(0,1){12}}

\put(80,18){\line(0,1){12}} \put(80,20){\circle{4}}
\put(80,30){\circle*{2}}

\end{picture}
\vskip2em\caption{Quantum {\small{SWAP}} gate for qutrit
states.}\label{effswap}
\end{figure}
Fig. \ref{effswap} gives a  generalised quantum {\small{SWAP}}
gate for qutrit states which is based upon our
binomial summation construction. By Theorem 2.4 
(Wilmott and Wild (2008c)), let us suppose that the first quantum
system ${\mathcal{A}}_0$ prepared in the state $\qudit{e_0}_0$,
the second system ${\mathcal{A}}_1$ prepared in the state
$\qudit{e_1}_1$ and  the third system ${\mathcal{A}}_2$ prepared
in the state $\qudit{e_2}_2$. Implementing our  quantum
{\small{SWAP}} gate yields the system ${\mathcal{A}}_0$ in the
state $\qudit{e_1}_0$, the  system ${\mathcal{A}}_1$ is in the
state $\qudit{e_2}_1$ and the system ${\mathcal{A}}_{{2}}$ is in
the state $\qudit{e_0}_2$ and this is achieved with eight
{\small{CNOT}} gates as opposed to the  ten {\small{CNOT}} gates
of the WilNOT construction. Note further that our  construction is
based on a set of binomial coefficients which satisfy the linear
recurrence relation whose generating  function is given by
$1/(1-z-z^{p^m})$. Generating functions provide elegant means for
the storing  of information about the nature of coefficients in
their expansion (Blasiak \emph{et al.} (2007)). For our purpose,
the nature of the recurrence relation is seen through periodicity
of the coefficients modulo 3. We relate this periodic nature to
instances of the {\small{CNOT}} gate acting on basis states in
determination of a quantum swap gate. Furthermore, since our
quantum gate is based on a linear recurrence relation, we
illustrate that knowing part of the quantum gate design will
completely aid in determining the remainder of the quantum gate
just as knowing the initial conditions of a recurrence relation
determines all other coefficients in the recurrence. Such type of
construction is unusual in that knowing part of the gate does not
necessarily yield the remainder.

We illustrate the action of each use of the generalised
{\small{CNOT}} gate arising in our  construction on the input
quantum states and subsequent evolved states of the system. The
first unitary gate $U_1$ acts on the input state of three
arbitrary qutrits where the first quantum system ${\mathcal{A}}_0$
on the input state is  prepared in the state $\qudit{e_0}_0$, the
second system ${\mathcal{A}}_1$ on the input state is prepared in
the state $\qudit{e_1}_1$ and  the third system ${\mathcal{A}}_2$
on the input state is prepared in the state $\qudit{e_2}_2$. The
action of $U_1$ in the input state is given by

\begin{eqnarray}
&& U_1(\qudit{a}\otimes\qudit{b}\otimes\qudit{c})\nonumber\\
&& = U_1\left((a_0\qudit{0}+a_1\qudit{1}+a_2\qudit{2})\otimes (b_0\qudit{0}+b_1
\qudit{1}+b_2\qudit{2})\otimes (c_0\qudit{0}+c_1\qudit{1}+c_2\qudit{2})\right) \nonumber\\
&&=
U_1(a_0b_0c_0\qudit{000}+a_0b_0c_1\qudit{001}+a_0b_0c_2\qudit{002}+a_0b_1c_0\qudit{010}+
a_0b_1c_1\qudit{011}\nonumber\\&&
\hskip1em+a_0b_1c_2\qudit{012}+a_0b_2c_0\qudit{020}+a_0b_2c_1\qudit{021}+a_0b_2c_2\qudit{022}+a_1b_0c_0\qudit{100}
\nonumber\\
&&\hskip1em+a_1b_0c_1\qudit{101}+a_1b_0c_2\qudit{102}+a_1b_1c_0\qudit{110}
+a_1b_1c_1\qudit{111}+a_1b_1c_2\qudit{112}\nonumber\\&&\hskip1em+a_1b_2c_0\qudit{120}+a_1b_2c_1
\qudit{121}+a_1b_2c_2\qudit{122}+a_2b_0c_0\qudit{200}+a_2b_0c_1\qudit{201}\nonumber\\&&\hskip1em
+a_2b_0c_2\qudit{202}+a_2b_1c_0\qudit{210}+a_2b_1c_1\qudit{211}+a_2b_1c_2\qudit{212}+a_2b_2c_0
\qudit{220}\nonumber\\&&\hskip1em+a_2b_2c_1\qudit{221}+a_2b_2c_2\qudit{222})
\nonumber\end{eqnarray}\begin{eqnarray}&& = a_0b_0c_0
\qudit{000}+a_0b_0c_1\qudit{001}+a_0b_0c_2\qudit{002}+a_0b_1c_0\qudit{010}+a_0b_1c_1\qudit{011}
\nonumber\\&&\hskip1em+a_0b_1c_2\qudit{012}+a_0b_2c_0\qudit{020}+a_0b_2c_1\qudit{021}+a_0b_2c_2\qudit{022}
+a_1b_0c_0\qudit{110}\nonumber\\&&\hskip1em+a_1b_0c_1\qudit{111}+a_1b_0c_2\qudit{112}+a_1b_1c_0\qudit{120}+
a_1b_1c_1\qudit{121}+a_1b_1c_2\qudit{122}\nonumber\\&&\hskip1em+a_1b_2c_0\qudit{100}+a_1b_2c_1\qudit{101}+
a_1b_2c_2\qudit{102}+a_2b_0c_0\qudit{220}+a_2b_0c_1\qudit{221}\nonumber\\&&\hskip1em+a_2b_0c_2\qudit{222}+
a_2b_1c_0\qudit{200}+a_2b_1c_1\qudit{201}+a_2b_1c_2\qudit{202}+a_2b_2c_0\qudit{210}\nonumber\\&&\hskip1em+
a_2b_2c_1\qudit{211}+a_2b_2c_2\qudit{212}.
\end{eqnarray}
\begin{figure} $\left(\begin{matrix} {1\ 0\
0\ 0\ 0\ 0\ 0\ 0\ 0\ 0\ 0\ 0\ 0\ 0\ 0\ 0\ 0\ 0\ 0\ 0\ 0\ 0\ 0\ 0\
0\ 0\ 0}\cr {0\ 1\ 0\ 0\ 0\ 0\ 0\ 0\ 0\ 0\ 0\ 0\ 0\ 0\ 0\ 0\ 0\ 0\
0\ 0\ 0\ 0\ 0\ 0\ 0\ 0\ 0}\cr {0\ 0\ 1\ 0\ 0\ 0\ 0\ 0\ 0\ 0\ 0\ 0\
0\ 0\ 0\ 0\ 0\ 0\ 0\ 0\ 0\ 0\ 0\ 0\ 0\ 0\ 0}\cr {0\ 0\ 0\ 1\ 0\ 0\
0\ 0\ 0\ 0\ 0\ 0\ 0\ 0\ 0\ 0\ 0\ 0\ 0\ 0\ 0\ 0\ 0\ 0\ 0\ 0\ 0}\cr
{0\ 0\ 0\ 0\ 1\ 0\ 0\ 0\ 0\ 0\ 0\ 0\ 0\ 0\ 0\ 0\ 0\ 0\ 0\ 0\ 0\ 0\
0\ 0\ 0\ 0\ 0}\cr {0\ 0\ 0\ 0\ 0\ 1\ 0\ 0\ 0\ 0\ 0\ 0\ 0\ 0\ 0\ 0\
0\ 0\ 0\ 0\ 0\ 0\ 0\ 0\ 0\ 0\ 0}\cr {0\ 0\ 0\ 0\ 0\ 0\ 1\ 0\ 0\ 0\
0\ 0\ 0\ 0\ 0\ 0\ 0\ 0\ 0\ 0\ 0\ 0\ 0\ 0\ 0\ 0\ 0}\cr {0\ 0\ 0\ 0\
0\ 0\ 0\ 1\ 0\ 0\ 0\ 0\ 0\ 0\ 0\ 0\ 0\ 0\ 0\ 0\ 0\ 0\ 0\ 0\ 0\ 0\
0}\cr {0\ 0\ 0\ 0\ 0\ 0\ 0\ 0\ 1\ 0\ 0\ 0\ 0\ 0\ 0\ 0\ 0\ 0\ 0\ 0\
0\ 0\ 0\ 0\ 0\ 0\ 0}\cr {0\ 0\ 0\ 0\ 0\ 0\ 0\ 0\ 0\ 0\ 0\ 0\ 0\ 0\
0\ 1\ 0\ 0\ 0\ 0\ 0\ 0\ 0\ 0\ 0\ 0\ 0}\cr {0\ 0\ 0\ 0\ 0\ 0\ 0\ 0\
0\ 0\ 0\ 0\ 0\ 0\ 0\ 0\ 1\ 0\ 0\ 0\ 0\ 0\ 0\ 0\ 0\ 0\ 0}\cr {0\ 0\
0\ 0\ 0\ 0\ 0\ 0\ 0\ 0\ 0\ 0\ 0\ 0\ 0\ 0\ 0\ 1\ 0\ 0\ 0\ 0\ 0\ 0\
0\ 0\ 0}\cr {0\ 0\ 0\ 0\ 0\ 0\ 0\ 0\ 0\ 1\ 0\ 0\ 0\ 0\ 0\ 0\ 0\ 0\
0\ 0\ 0\ 0\ 0\ 0\ 0\ 0\ 0}\cr {0\ 0\ 0\ 0\ 0\ 0\ 0\ 0\ 0\ 0\ 1\ 0\
0\ 0\ 0\ 0\ 0\ 0\ 0\ 0\ 0\ 0\ 0\ 0\ 0\ 0\ 0}\cr {0\ 0\ 0\ 0\ 0\ 0\
0\ 0\ 0\ 0\ 0\ 1\ 0\ 0\ 0\ 0\ 0\ 0\ 0\ 0\ 0\ 0\ 0\ 0\ 0\ 0\ 0}\cr
{0\ 0\ 0\ 0\ 0\ 0\ 0\ 0\ 0\ 0\ 0\ 0\ 1\ 0\ 0\ 0\ 0\ 0\ 0\ 0\ 0\ 0\
0\ 0\ 0\ 0\ 0}\cr {0\ 0\ 0\ 0\ 0\ 0\ 0\ 0\ 0\ 0\ 0\ 0\ 0\ 1\ 0\ 0\
0\ 0\ 0\ 0\ 0\ 0\ 0\ 0\ 0\ 0\ 0}\cr {0\ 0\ 0\ 0\ 0\ 0\ 0\ 0\ 0\ 0\
0\ 0\ 0\ 0\ 1\ 0\ 0\ 0\ 0\ 0\ 0\ 0\ 0\ 0\ 0\ 0\ 0}\cr {0\ 0\ 0\ 0\
0\ 0\ 0\ 0\ 0\ 0\ 0\ 0\ 0\ 0\ 0\ 0\ 0\ 0\ 0\ 0\ 0\ 1\ 0\ 0\ 0\ 0\
0}\cr {0\ 0\ 0\ 0\ 0\ 0\ 0\ 0\ 0\ 0\ 0\ 0\ 0\ 0\ 0\ 0\ 0\ 0\ 0\ 0\
0\ 0\ 1\ 0\ 0\ 0\ 0}\cr {0\ 0\ 0\ 0\ 0\ 0\ 0\ 0\ 0\ 0\ 0\ 0\ 0\ 0\
0\ 0\ 0\ 0\ 0\ 0\ 0\ 0\ 0\ 1\ 0\ 0\ 0}\cr {0\ 0\ 0\ 0\ 0\ 0\ 0\ 0\
0\ 0\ 0\ 0\ 0\ 0\ 0\ 0\ 0\ 0\ 0\ 0\ 0\ 0\ 0\ 0\ 1\ 0\ 0}\cr {0\ 0\
0\ 0\ 0\ 0\ 0\ 0\ 0\ 0\ 0\ 0\ 0\ 0\ 0\ 0\ 0\ 0\ 0\ 0\ 0\ 0\ 0\ 0\
0\ 1\ 0}\cr {0\ 0\ 0\ 0\ 0\ 0\ 0\ 0\ 0\ 0\ 0\ 0\ 0\ 0\ 0\ 0\ 0\ 0\
0\ 0\ 0\ 0\ 0\ 0\ 0\ 0\ 1}\cr {0\ 0\ 0\ 0\ 0\ 0\ 0\ 0\ 0\ 0\ 0\ 0\
0\ 0\ 0\ 0\ 0\ 0\ 1\ 0\ 0\ 0\ 0\ 0\ 0\ 0\ 0}\cr {0\ 0\ 0\ 0\ 0\ 0\
0\ 0\ 0\ 0\ 0\ 0\ 0\ 0\ 0\ 0\ 0\ 0\ 0\ 1\ 0\ 0\ 0\ 0\ 0\ 0\ 0}\cr
{0\ 0\ 0\ 0\ 0\ 0\ 0\ 0\ 0\ 0\ 0\ 0\ 0\ 0\ 0\ 0\ 0\ 0\ 0\ 0\ 1\ 0\
0\ 0\ 0\ 0\ 0}\cr
\end{matrix}
\right)$ \caption{The unitary matrix $U_1$}
\end{figure}
\noindent The second unitary matrix $U_2$ corresponds to the
second {\small{CNOT}} gate given in Fig.~\ref{effswap} and  acts
on the state of the system after application of the first unitary
transformation. The action of the unitary matrix $U_2$ on the
state of the system
\begin{eqnarray}
&&U_1(a_0b_0c_0\qudit{000}+a_0b_0c_1\qudit{001}+a_0b_0c_2\qudit{002}+a_0b_1c_0\qudit{010}+a_0b_1c_1\qudit{011}\nonumber
\\&&\hskip1em+a_0b_1c_2\qudit{012}+a_0b_2c_0\qudit{020}+a_0b_2c_1\qudit{021}+a_0b_2c_2\qudit{022}+a_1b_0c_0
\qudit{110}\nonumber\\&&\hskip1em+a_1b_0c_1\qudit{111}+a_1b_0c_2\qudit{112}+a_1b_1c_0
\qudit{120}+a_1b_1c_1\qudit{121}+a_1b_1c_2\qudit{122}\nonumber\\&&\hskip1em+a_1b_2c_0\qudit{100}
+a_1b_2c_1\qudit{101}+a_1b_2c_2\qudit{102}+a_2b_0c_0\qudit{220}+a_2b_0c_1\qudit{221}\nonumber\\&&\hskip1em
+a_2b_0c_2\qudit{222}+a_2b_1c_0\qudit{200}+a_2b_1c_1\qudit{201}+a_2b_1c_2
\qudit{202}+a_2b_2c_0\qudit{210}\nonumber\\&&\hskip1em+a_2b_2c_1\qudit{211}+a_2b_2c_2\qudit{212})\nonumber\end{eqnarray}
is
\begin{figure}
$\left(\begin{matrix} {1\ 0\ 0\ 0\ 0\ 0\ 0\ 0\ 0\ 0\ 0\ 0\ 0\ 0\
0\ 0\ 0\ 0\ 0\ 0\ 0\ 0\ 0\ 0\ 0\ 0\ 0}\cr {0\ 1\ 0\ 0\ 0\ 0\ 0\ 0\
0\ 0\ 0\ 0\ 0\ 0\ 0\ 0\ 0\ 0\ 0\ 0\ 0\ 0\ 0\ 0\ 0\ 0\ 0}\cr {0\ 0\
1\ 0\ 0\ 0\ 0\ 0\ 0\ 0\ 0\ 0\ 0\ 0\ 0\ 0\ 0\ 0\ 0\ 0\ 0\ 0\ 0\ 0\
0\ 0\ 0}\cr {0\ 0\ 0\ 0\ 0\ 1\ 0\ 0\ 0\ 0\ 0\ 0\ 0\ 0\ 0\ 0\ 0\ 0\
0\ 0\ 0\ 0\ 0\ 0\ 0\ 0\ 0}\cr {0\ 0\ 0\ 1\ 0\ 0\ 0\ 0\ 0\ 0\ 0\ 0\
0\ 0\ 0\ 0\ 0\ 0\ 0\ 0\ 0\ 0\ 0\ 0\ 0\ 0\ 0}\cr {0\ 0\ 0\ 0\ 1\ 0\
0\ 0\ 0\ 0\ 0\ 0\ 0\ 0\ 0\ 0\ 0\ 0\ 0\ 0\ 0\ 0\ 0\ 0\ 0\ 0\ 0}\cr
{0\ 0\ 0\ 0\ 0\ 0\ 0\ 1\ 0\ 0\ 0\ 0\ 0\ 0\ 0\ 0\ 0\ 0\ 0\ 0\ 0\ 0\
0\ 0\ 0\ 0\ 0}\cr {0\ 0\ 0\ 0\ 0\ 0\ 0\ 0\ 1\ 0\ 0\ 0\ 0\ 0\ 0\ 0\
0\ 0\ 0\ 0\ 0\ 0\ 0\ 0\ 0\ 0\ 0}\cr {0\ 0\ 0\ 0\ 0\ 0\ 1\ 0\ 0\ 0\
0\ 0\ 0\ 0\ 0\ 0\ 0\ 0\ 0\ 0\ 0\ 0\ 0\ 0\ 0\ 0\ 0}\cr {0\ 0\ 0\ 0\
0\ 0\ 0\ 0\ 0\ 1\ 0\ 0\ 0\ 0\ 0\ 0\ 0\ 0\ 0\ 0\ 0\ 0\ 0\ 0\ 0\ 0\
0}\cr {0\ 0\ 0\ 0\ 0\ 0\ 0\ 0\ 0\ 0\ 1\ 0\ 0\ 0\ 0\ 0\ 0\ 0\ 0\ 0\
0\ 0\ 0\ 0\ 0\ 0\ 0}\cr {0\ 0\ 0\ 0\ 0\ 0\ 0\ 0\ 0\ 0\ 0\ 1\ 0\ 0\
0\ 0\ 0\ 0\ 0\ 0\ 0\ 0\ 0\ 0\ 0\ 0\ 0}\cr {0\ 0\ 0\ 0\ 0\ 0\ 0\ 0\
0\ 0\ 0\ 0\ 0\ 0\ 1\ 0\ 0\ 0\ 0\ 0\ 0\ 0\ 0\ 0\ 0\ 0\ 0}\cr {0\ 0\
0\ 0\ 0\ 0\ 0\ 0\ 0\ 0\ 0\ 0\ 1\ 0\ 0\ 0\ 0\ 0\ 0\ 0\ 0\ 0\ 0\ 0\
0\ 0\ 0}\cr {0\ 0\ 0\ 0\ 0\ 0\ 0\ 0\ 0\ 0\ 0\ 0\ 0\ 1\ 0\ 0\ 0\ 0\
0\ 0\ 0\ 0\ 0\ 0\ 0\ 0\ 0}\cr {0\ 0\ 0\ 0\ 0\ 0\ 0\ 0\ 0\ 0\ 0\ 0\
0\ 0\ 0\ 0\ 1\ 0\ 0\ 0\ 0\ 0\ 0\ 0\ 0\ 0\ 0}\cr {0\ 0\ 0\ 0\ 0\ 0\
0\ 0\ 0\ 0\ 0\ 0\ 0\ 0\ 0\ 0\ 0\ 1\ 0\ 0\ 0\ 0\ 0\ 0\ 0\ 0\ 0}\cr
{0\ 0\ 0\ 0\ 0\ 0\ 0\ 0\ 0\ 0\ 0\ 0\ 0\ 0\ 0\ 1\ 0\ 0\ 0\ 0\ 0\ 0\
0\ 0\ 0\ 0\ 0}\cr {0\ 0\ 0\ 0\ 0\ 0\ 0\ 0\ 0\ 0\ 0\ 0\ 0\ 0\ 0\ 0\
0\ 0\ 1\ 0\ 0\ 0\ 0\ 0\ 0\ 0\ 0}\cr {0\ 0\ 0\ 0\ 0\ 0\ 0\ 0\ 0\ 0\
0\ 0\ 0\ 0\ 0\ 0\ 0\ 0\ 0\ 1\ 0\ 0\ 0\ 0\ 0\ 0\ 0}\cr {0\ 0\ 0\ 0\
0\ 0\ 0\ 0\ 0\ 0\ 0\ 0\ 0\ 0\ 0\ 0\ 0\ 0\ 0\ 0\ 1\ 0\ 0\ 0\ 0\ 0\
0}\cr {0\ 0\ 0\ 0\ 0\ 0\ 0\ 0\ 0\ 0\ 0\ 0\ 0\ 0\ 0\ 0\ 0\ 0\ 0\ 0\
0\ 0\ 0\ 1\ 0\ 0\ 0}\cr {0\ 0\ 0\ 0\ 0\ 0\ 0\ 0\ 0\ 0\ 0\ 0\ 0\ 0\
0\ 0\ 0\ 0\ 0\ 0\ 0\ 1\ 0\ 0\ 0\ 0\ 0}\cr {0\ 0\ 0\ 0\ 0\ 0\ 0\ 0\
0\ 0\ 0\ 0\ 0\ 0\ 0\ 0\ 0\ 0\ 0\ 0\ 0\ 0\ 1\ 0\ 0\ 0\ 0}\cr {0\ 0\
0\ 0\ 0\ 0\ 0\ 0\ 0\ 0\ 0\ 0\ 0\ 0\ 0\ 0\ 0\ 0\ 0\ 0\ 0\ 0\ 0\ 0\
0\ 1\ 0}\cr {0\ 0\ 0\ 0\ 0\ 0\ 0\ 0\ 0\ 0\ 0\ 0\ 0\ 0\ 0\ 0\ 0\ 0\
0\ 0\ 0\ 0\ 0\ 0\ 0\ 0\ 1}\cr {0\ 0\ 0\ 0\ 0\ 0\ 0\ 0\ 0\ 0\ 0\ 0\
0\ 0\ 0\ 0\ 0\ 0\ 0\ 0\ 0\ 0\ 0\ 0\ 1\ 0\ 0}\cr
\end{matrix}\right)$
\caption{The unitary matrix $U_2$}
\end{figure}
\begin{eqnarray}
&&U_2(a_0b_0c_0\qudit{000}+a_0b_0c_1\qudit{001}+a_0b_0c_2\qudit{002}+a_0b_1c_0\qudit{010}
+a_0b_1c_1\qudit{011}\nonumber\\&&\hskip1em+a_0b_1c_2\qudit{012}+a_0b_2c_0\qudit{020}
+a_0b_2c_1\qudit{021}+a_0b_2c_2\qudit{022}+a_1b_0c_0\qudit{110}\nonumber\\&&\hskip1em+a_1b_0c_1
\qudit{111}+a_1b_0c_2\qudit{112}+a_1b_1c_0\qudit{120}+a_1b_1c_1\qudit{121}+a_1b_1c_2\qudit{122}\nonumber\\&&
\hskip1em+a_1b_2c_0\qudit{100}+a_1b_2c_1\qudit{101}+a_1b_2c_2\qudit{102}+a_2b_0c_0\qudit{220}
+a_2b_0c_1\qudit{221}\nonumber\\&&\hskip1em+a_2b_0c_2\qudit{222}+a_2b_1c_0\qudit{200}+a_2b_1c_1\qudit{201}
+a_2b_1c_2\qudit{202}+a_2b_2c_0\qudit{210}\nonumber\\&&\hskip1em+a_2b_2c_1\qudit{211}+a_2b_2c_2\qudit{212})\nonumber\\&&
= a_0b_0c_0\qudit{000}+a_0b_0c_1\qudit{001}+a_0b_0c_2\qudit{002}
+a_0b_1c_0\qudit{011}+a_0b_1c_1\qudit{012}\nonumber\\&&\hskip1em+a_0b_1c_2\qudit{010}+
a_0b_2c_0\qudit{022}+a_0b_2c_1\qudit{020}+a_0b_2c_2\qudit{021}+a_1b_0c_0\qudit{111}\nonumber\\&&\hskip1em
+a_1b_0c_1\qudit{112}+a_1b_0c_2\qudit{110}+a_1b_1c_0\qudit{122}+a_1b_1c_1\qudit{120}+a_1b_1c_2\qudit{121}\nonumber\\&&\hskip1em
+a_1b_2c_0\qudit{100}+a_1b_2c_1\qudit{101}+a_1b_2c_2\qudit{102}+a_2b_0c_0\qudit{222}+a_2b_0c_1\qudit{220}\nonumber\\&&\hskip1em
+a_2b_0c_2\qudit{221}+a_2b_1c_0\qudit{200}+a_2b_1c_1\qudit{201}+a_2b_1c_2\qudit{202}+a_2b_2c_0\qudit{211}\nonumber\\&&\hskip1em
+a_2b_2c_1\qudit{212}+a_2b_2c_2\qudit{210}.
\end{eqnarray}

\noindent  The unitary matrix $U_3$ is given in Fig.~\ref{333} and
\begin{figure}
$\left(\begin{matrix} {1\ 0\ 0\ 0\ 0\ 0\ 0\ 0\ 0\ 0\ 0\ 0\ 0\ 0\
0\ 0\ 0\ 0\ 0\ 0\ 0\ 0\ 0\ 0\ 0\ 0\ 0}\cr {0\ 0\ 0\ 0\ 0\ 0\ 0\ 0\
0\ 0\ 0\ 0\ 0\ 0\ 0\ 0\ 0\ 0\ 0\ 1\ 0\ 0\ 0\ 0\ 0\ 0\ 0}\cr {0\ 0\
0\ 0\ 0\ 0\ 0\ 0\ 0\ 0\ 0\ 1\ 0\ 0\ 0\ 0\ 0\ 0\ 0\ 0\ 0\ 0\ 0\ 0\
0\ 0\ 0}\cr {0\ 0\ 0\ 1\ 0\ 0\ 0\ 0\ 0\ 0\ 0\ 0\ 0\ 0\ 0\ 0\ 0\ 0\
0\ 0\ 0\ 0\ 0\ 0\ 0\ 0\ 0}\cr {0\ 0\ 0\ 0\ 0\ 0\ 0\ 0\ 0\ 0\ 0\ 0\
0\ 0\ 0\ 0\ 0\ 0\ 0\ 0\ 0\ 0\ 1\ 0\ 1\ 0\ 0}\cr {0\ 0\ 0\ 0\ 0\ 0\
0\ 0\ 0\ 0\ 0\ 0\ 0\ 0\ 1\ 0\ 0\ 0\ 0\ 0\ 0\ 0\ 0\ 0\ 0\ 0\ 0}\cr
{0\ 0\ 0\ 0\ 0\ 0\ 1\ 0\ 0\ 0\ 0\ 0\ 0\ 0\ 0\ 0\ 0\ 0\ 0\ 0\ 0\ 0\
0\ 0\ 0\ 0\ 0}\cr {0\ 0\ 0\ 0\ 0\ 0\ 0\ 0\ 0\ 0\ 0\ 0\ 0\ 0\ 0\ 0\
0\ 0\ 0\ 0\ 0\ 0\ 0\ 0\ 0\ 1\ 0}\cr {0\ 0\ 0\ 0\ 0\ 0\ 0\ 0\ 0\ 0\
0\ 0\ 0\ 0\ 0\ 0\ 0\ 1\ 0\ 0\ 0\ 0\ 0\ 0\ 0\ 0\ 0}\cr {0\ 0\ 0\ 0\
0\ 0\ 0\ 0\ 0\ 1\ 0\ 0\ 0\ 0\ 0\ 0\ 0\ 0\ 0\ 0\ 0\ 0\ 0\ 0\ 0\ 0\
0}\cr {0\ 0\ 0\ 0\ 0\ 0\ 0\ 0\ 0\ 0\ 0\ 0\ 0\ 0\ 0\ 0\ 0\ 0\ 0\ 0\
0\ 0\ 0\ 0\ 0\ 0\ 0}\cr {0\ 0\ 1\ 0\ 0\ 0\ 0\ 0\ 0\ 0\ 0\ 0\ 0\ 0\
0\ 0\ 0\ 0\ 0\ 0\ 1\ 0\ 0\ 0\ 0\ 0\ 0}\cr {0\ 0\ 0\ 0\ 0\ 0\ 0\ 0\
0\ 0\ 0\ 0\ 1\ 0\ 0\ 0\ 0\ 0\ 0\ 0\ 0\ 0\ 0\ 0\ 0\ 0\ 0}\cr {0\ 0\
0\ 0\ 1\ 0\ 0\ 0\ 0\ 0\ 0\ 0\ 0\ 0\ 0\ 0\ 0\ 0\ 0\ 0\ 0\ 0\ 0\ 0\
0\ 0\ 0}\cr {0\ 0\ 0\ 0\ 0\ 0\ 0\ 0\ 0\ 0\ 0\ 0\ 0\ 0\ 0\ 0\ 0\ 0\
0\ 0\ 0\ 0\ 0\ 1\ 0\ 0\ 0}\cr {0\ 0\ 0\ 0\ 0\ 0\ 0\ 0\ 0\ 0\ 0\ 0\
0\ 0\ 0\ 1\ 0\ 0\ 0\ 0\ 0\ 0\ 0\ 0\ 0\ 0\ 0}\cr {0\ 0\ 0\ 0\ 0\ 0\
0\ 1\ 0\ 0\ 0\ 0\ 0\ 0\ 0\ 0\ 0\ 0\ 0\ 0\ 0\ 0\ 0\ 0\ 0\ 0\ 0}\cr
{0\ 0\ 0\ 0\ 0\ 0\ 0\ 0\ 0\ 0\ 0\ 0\ 0\ 0\ 0\ 0\ 0\ 0\ 0\ 0\ 0\ 0\
0\ 0\ 0\ 0\ 1}\cr {0\ 0\ 0\ 0\ 0\ 0\ 0\ 0\ 0\ 0\ 0\ 0\ 0\ 0\ 0\ 0\
0\ 0\ 1\ 0\ 0\ 0\ 0\ 0\ 0\ 0\ 0}\cr {0\ 1\ 0\ 0\ 0\ 0\ 0\ 0\ 0\ 0\
1\ 0\ 0\ 0\ 0\ 0\ 0\ 0\ 0\ 0\ 0\ 0\ 0\ 0\ 0\ 0\ 0}\cr {0\ 0\ 0\ 0\
0\ 0\ 0\ 0\ 0\ 0\ 0\ 0\ 0\ 0\ 0\ 0\ 0\ 0\ 0\ 0\ 0\ 0\ 0\ 0\ 0\ 0\
0}\cr {0\ 0\ 0\ 0\ 0\ 0\ 0\ 0\ 0\ 0\ 0\ 0\ 0\ 0\ 0\ 0\ 0\ 0\ 0\ 0\
0\ 1\ 0\ 0\ 0\ 0\ 0}\cr {0\ 0\ 0\ 0\ 0\ 0\ 0\ 0\ 0\ 0\ 0\ 0\ 0\ 1\
0\ 0\ 0\ 0\ 0\ 0\ 0\ 0\ 0\ 0\ 0\ 0\ 0}\cr {0\ 0\ 0\ 0\ 0\ 1\ 0\ 0\
0\ 0\ 0\ 0\ 0\ 0\ 0\ 0\ 0\ 0\ 0\ 0\ 0\ 0\ 0\ 0\ 0\ 0\ 0}\cr {0\ 0\
0\ 0\ 0\ 0\ 0\ 0\ 0\ 0\ 0\ 0\ 0\ 0\ 0\ 0\ 0\ 0\ 0\ 0\ 0\ 0\ 0\ 0\
1\ 0\ 0}\cr {0\ 0\ 0\ 0\ 0\ 0\ 0\ 0\ 0\ 0\ 0\ 0\ 0\ 0\ 0\ 0\ 1\ 0\
0\ 0\ 0\ 0\ 0\ 0\ 0\ 0\ 0}\cr {0\ 0\ 0\ 0\ 0\ 0\ 0\ 0\ 1\ 0\ 0\ 0\
0\ 0\ 0\ 0\ 0\ 0\ 0\ 0\ 0\ 0\ 0\ 0\ 0\ 0\
0}\cr\end{matrix}\right)$ \caption{The unitary matrix
$U_3$}\label{333}
\end{figure}
\noindent corresponds to the third {\small{CNOT}} gate which acts
on the state of the system after application of the second
unitary. Its action on the state
\begin{eqnarray}
&&
U_2U_1(a_0b_0c_0\qudit{000}+a_0b_0c_1\qudit{001}+a_0b_0c_2\qudit{002}+a_0b_1c_0
\qudit{010}+a_0b_1c_1\qudit{011}\nonumber\\&&\hskip1em
+a_0b_1c_2\qudit{012}+a_0b_2c_0\qudit{020}+a_0b_2c_1\qudit{021}+a_0b_2c_2\qudit{022}+a_1b_0c_0\qudit{110}\nonumber\\&&\hskip1em
+a_1b_0c_1\qudit{111}+a_1b_0c_2\qudit{112}+a_1b_1c_0\qudit{120}+a_1b_1c_1\qudit{121}+a_1b_1c_2\qudit{122}\nonumber\\&&\hskip1em
+a_1b_2c_0\qudit{100}+a_1b_2c_1\qudit{101}+a_1b_2c_2\qudit{102}+a_2b_0c_0\qudit{220}+a_2b_0c_1\qudit{221}\nonumber\\&&\hskip1em
+a_2b_0c_2\qudit{222}+a_2b_1c_0\qudit{200}+a_2b_1c_1\qudit{201}+a_2b_1c_2\qudit{202}+a_2b_2c_0\qudit{210}\nonumber\\&&\hskip1em
+a_2b_2c_1\qudit{211}+a_2b_2c_2\qudit{212})\nonumber
\end{eqnarray}
is
\begin{eqnarray}
&&U_3(a_0b_0c_0\qudit{000}+a_0b_0c_1\qudit{001}+a_0b_0c_2\qudit{002}+a_0b_1c_0\qudit{011}+a_0b_1c_1\qudit{012}
\nonumber\\&&\hskip1em
+a_0b_1c_2\qudit{010}+a_0b_2c_0\qudit{022}+a_0b_2c_1\qudit{020}+a_0b_2c_2\qudit{021}+a_1b_0c_0\qudit{111}\nonumber\\&&\hskip1em
+a_1b_0c_1\qudit{112}+a_1b_0c_2\qudit{110}+a_1b_1c_0\qudit{122}+a_1b_1c_1\qudit{120}+a_1b_1c_2\qudit{121}\nonumber\\&&\hskip1em
+a_1b_2c_0\qudit{100}+a_1b_2c_1\qudit{101}+a_1b_2c_2\qudit{102}+a_2b_0c_0\qudit{222}+a_2b_0c_1\qudit{220}\nonumber\\&&\hskip1em
+a_2b_0c_2\qudit{221}+a_2b_1c_0\qudit{200}+a_2b_1c_1\qudit{201}+a_2b_1c_2\qudit{202}+a_2b_2c_0\qudit{211}\nonumber\\&&\hskip1em
+a_2b_2c_1\qudit{212}+a_2b_2c_2\qudit{210})
\nonumber\end{eqnarray}\begin{eqnarray} && =
a_0b_0c_0\qudit{000}+a_0b_0c_1\qudit{101}+a_0b_0c_2\qudit{202}+a_0b_1c_0\qudit{111}+a_0b_1c_1\qudit{212}\nonumber\\&&\hskip1em
+a_0b_1c_2\qudit{010}+a_0b_2c_0\qudit{222}+a_0b_2c_1\qudit{020}+a_0b_2c_2\qudit{121}+a_1b_0c_0\qudit{211}\nonumber
\\&&\hskip1em
+a_1b_0c_1\qudit{012}+a_1b_0c_2\qudit{110}+a_1b_1c_0\qudit{022}+a_1b_1c_1\qudit{120}+a_1b_1c_2\qudit{221}\nonumber\\&&\hskip1em
+a_1b_2c_0\qudit{100}+a_1b_2c_1\qudit{201}+a_1b_2c_2\qudit{002}+a_2b_0c_0\qudit{122}+a_2b_0c_1\qudit{220}\nonumber\\&&\hskip1em
+a_2b_0c_2\qudit{021}+a_2b_1c_0\qudit{200}+a_2b_1c_1\qudit{001}+a_2b_1c_2\qudit{102}+a_2b_2c_0\qudit{011}\nonumber\\&&\hskip1em
+a_2b_2c_1\qudit{112}+a_2b_2c_2\qudit{210}.
\end{eqnarray}
As outlined, this construction is based on a linear recurrence
relation and therefore there is a cyclic nature to our gate
construction. This is evident when we consider our gate within the
qutrit setting. As such, our construction is based upon repeat
application of the unitrary matrices $U_1, U_2$, and $U_3$. As
evident in Fig. \ref{effswap} the fourth gate in the construction
corresponds to the unitary matrix $U_1$. In particular, we have it
that
\begin{eqnarray}
&&\hskip-3em
U_4(U_3U_2U_1(a_0b_0c_0\qudit{000}+a_0b_0c_1\qudit{001}+a_0b_0c_2\qudit{002}+a_0b_1c_0\
qudit{010}+a_0b_1c_1\qudit{011}\nonumber\\&&\hskip1em
+a_0b_1c_2\qudit{012}+a_0b_2c_0\qudit{020}+a_0b_2c_1\qudit{021}+a_0b_2c_2\qudit{022}+a_1b_0c_0\qudit{110}\nonumber\\&&\hskip1em
+a_1b_0c_1\qudit{111}+a_1b_0c_2\qudit{112}+a_1b_1c_0\qudit{120}+a_1b_1c_1\qudit{121}+a_1b_1c_2\qudit{122}\nonumber\\&&\hskip1em
+a_1b_2c_0\qudit{100}+a_1b_2c_1\qudit{101}+a_1b_2c_2\qudit{102}+a_2b_0c_0\qudit{220}+a_2b_0c_1\qudit{221}\nonumber\\&&\hskip1em
+a_2b_0c_2\qudit{222}+a_2b_1c_0\qudit{200}+a_2b_1c_1\qudit{201}+a_2b_1c_2\qudit{202}+a_2b_2c_0\qudit{210}\nonumber\\
&&\hskip1em
+a_2b_2c_1\qudit{211}+a_2b_2c_2\qudit{212}))\nonumber\\
&& =
U_4(a_0b_0c_0\qudit{000}+a_0b_0c_1\qudit{101}+a_0b_0c_2\qudit{202}+a_0b_1c_0\qudit{111}+
a_0b_1c_1\qudit{212}\nonumber\\&&\hskip1em
+a_0b_1c_2\qudit{010}+a_0b_2c_0\qudit{222}+a_0b_2c_1\qudit{020}+a_0b_2c_2\qudit{121}+a_1b_0c_0\qudit{211}\nonumber\\&&\hskip1em
+a_1b_0c_1\qudit{012}+a_1b_0c_2\qudit{110}+a_1b_1c_0\qudit{022}+a_1b_1c_1\qudit{120}+a_1b_1c_2\qudit{221}\nonumber\\&&\hskip1em
+a_1b_2c_0\qudit{100}+a_1b_2c_1\qudit{201}+a_1b_2c_2\qudit{002}+a_2b_0c_0\qudit{122}+a_2b_0c_1\qudit{220}\nonumber\\&&\hskip1em
+a_2b_0c_2\qudit{021}+a_2b_1c_0\qudit{200}+a_2b_1c_1\qudit{001}+a_2b_1c_2\qudit{102}+a_2b_2c_0\qudit{011}\nonumber\\&&\hskip1em
+a_2b_2c_1\qudit{112}+a_2b_2c_2\qudit{210})\nonumber\\
&& =
a_0b_0c_0\qudit{000}+a_0b_0c_1\qudit{111}+a_0b_0c_2\qudit{222}+a_0b_1c_0\qudit{121}+a_0b_1c_1
\qudit{202}\nonumber\\&&\hskip1em
+a_0b_1c_2\qudit{010}+a_0b_2c_0\qudit{212}+a_0b_2c_1\qudit{020}+a_0b_2c_2\qudit{101}+a_1b_0c_0\qudit{201}\nonumber\\&&\hskip1em
+a_1b_0c_1\qudit{012}+a_1b_0c_2\qudit{120}+a_1b_1c_0\qudit{022}+a_1b_1c_1\qudit{100}+a_1b_1c_2\qudit{211}\nonumber\\&&\hskip1em
+a_1b_2c_0\qudit{110}+a_1b_2c_1\qudit{221}+a_1b_2c_2\qudit{002}+a_2b_0c_0\qudit{102}+a_2b_0c_1\qudit{210}\nonumber\\&&\hskip1em
+a_2b_0c_2\qudit{021}+a_2b_1c_0\qudit{220}+a_2b_1c_1\qudit{001}+a_2b_1c_2\qudit{112}+a_2b_2c_0\qudit{011}\nonumber\\&&\hskip1em
+a_2b_2c_1\qudit{122}+a_2b_2c_2\qudit{200}.
\end{eqnarray}

\noindent In a similar fashion, we have it that
\begin{eqnarray}
&&U_5(U_4U_3U_2U_1(a_0b_0c_0\qudit{000}+a_0b_0c_1\qudit{001}+a_0b_0c_2\qudit{002}+a_0b_1c_0\qudit{010}\nonumber
\\&&\hskip1em
+a_0b_1c_1\qudit{011}+a_0b_1c_2\qudit{012}+a_0b_2c_0\qudit{020}+a_0b_2c_1\qudit{021}+a_0b_2c_2\qudit{022}\nonumber\\&&\hskip1em
+a_1b_0c_0\qudit{110}+a_1b_0c_1\qudit{111}+a_1b_0c_2\qudit{112}+a_1b_1c_0\qudit{120}+a_1b_1c_1\qudit{121}\nonumber\\&&\hskip1em
+a_1b_1c_2\qudit{122}+a_1b_2c_0\qudit{100}+a_1b_2c_1\qudit{101}+a_1b_2c_2\qudit{102}+a_2b_0c_0\qudit{220}\nonumber\\&&\hskip1em
+a_2b_0c_1\qudit{221}+a_2b_0c_2\qudit{222}+a_2b_1c_0\qudit{200}+a_2b_1c_1\qudit{201}+a_2b_1c_2\qudit{202}\nonumber\\&&\hskip1em
+a_2b_2c_0\qudit{210}+a_2b_2c_1\qudit{211}+a_2b_2c_2\qudit{212}))\nonumber\end{eqnarray}\begin{eqnarray}
&& =
U_5(a_0b_0c_0\qudit{000}+a_0b_0c_1\qudit{111}+a_0b_0c_2\qudit{222}+a_0b_1c_0\qudit{121}+a_0b_1c_1
\qudit{202}\nonumber\\&&\hskip1em+a_0b_1c_2\qudit{010}+a_0b_2c_0\qudit{212}+a_0b_2c_1\qudit{020}+a_0b_2c_2\qudit{101}
+a_1b_0c_0\qudit{201}\nonumber\\&&\hskip1em
+a_1b_0c_1\qudit{012}+a_1b_0c_2\qudit{120}+a_1b_1c_0\qudit{022}+a_1b_1c_1\qudit{100}+a_1b_1c_2\qudit{211}\nonumber\\&&\hskip1em
+a_1b_2c_0\qudit{110}+a_1b_2c_1\qudit{221}+a_1b_2c_2\qudit{002}+a_2b_0c_0\qudit{102}+a_2b_0c_1\qudit{210}\nonumber\\&&\hskip1em
+a_2b_0c_2\qudit{021}+a_2b_1c_0\qudit{220}+a_2b_1c_1\qudit{001}+a_2b_1c_2\qudit{112}+a_2b_2c_0\qudit{011}\nonumber\\&&\hskip1em
+a_2b_2c_1\qudit{122}+a_2b_2c_2\qudit{200}) \nonumber\\&& =
a_0b_0c_0\qudit{000}+a_0b_0c_1\qudit{112}+a_0b_0c_2\qudit{221}+a_0b_1c_0\qudit{120}+a_0b_1c_1\qudit{202}\nonumber\\&&\hskip1em
+a_0b_1c_2\qudit{011}+a_0b_2c_0\qudit{210}+a_0b_2c_1\qudit{022}+a_0b_2c_2\qudit{101}+a_1b_0c_0\qudit{201}\nonumber\\&&\hskip1em
+a_1b_0c_1\qudit{010}+a_1b_0c_2\qudit{122}+a_1b_1c_0\qudit{021}+a_1b_1c_1\qudit{100}+a_1b_1c_2\qudit{212}\nonumber\\&&\hskip1em
+a_1b_2c_0\qudit{111}+a_1b_2c_1\qudit{220}+a_1b_2c_2\qudit{002}+a_2b_0c_0\qudit{102}+a_2b_0c_1\qudit{211}\nonumber\\&&\hskip1em
+a_2b_0c_2\qudit{020}+a_2b_1c_0\qudit{222}+a_2b_1c_1\qudit{001}+a_2b_1c_2\qudit{110}+a_2b_2c_0\qudit{012}\nonumber\\&&\hskip1em
+a_2b_2c_1\qudit{121}+a_2b_2c_2\qudit{200}
\end{eqnarray}
where the unitary matrix $U_5$ is given by the matrix $U_2$. The
next unitary matrix $U_6$, and  as illustrated in Fig.
\ref{effswap}, given by this construction is the matrix which
corresponds to the matrix $U_3$. We then  have it that
\begin{eqnarray}
&&U_6(U_5U_4U_3U_2U_1(a_0b_0c_0\qudit{000}+a_0b_0c_1\qudit{001}+a_0b_0c_2\qudit{002}+a_0b_1c_0\qudit{010}\nonumber
\\&&\hskip1em+a_0b_1c_1\qudit{011}+a_0b_1c_2\qudit{012}+a_0b_2c_0\qudit{020}+a_0b_2c_1\qudit{021}+a_0b_2c_2\qudit{022}\nonumber
\\&&\hskip1em+a_1b_0c_0\qudit{110}+a_1b_0c_1\qudit{111}+a_1b_0c_2\qudit{112}+a_1b_1c_0\qudit{120}+a_1b_1c_1\qudit{121}
\nonumber\\&&\hskip1em+a_1b_1c_2\qudit{122}+a_1b_2c_0
\qudit{100}+a_1b_2c_1\qudit{101}+a_1b_2c_2\qudit{102}+a_2b_0c_0\qudit{220}\nonumber\\&&\hskip1em
+a_2b_0c_1\qudit{221}+a_2b_0c_2\qudit{222}+a_2b_1c_0\qudit{200}+a_2b_1c_1\qudit{201}+a_2b_1c_2\qudit{202}\nonumber\\&&\hskip1em
+a_2b_2c_0\qudit{210}+a_2b_2c_1\qudit{211}+a_2b_2c_2\qudit{212}))\nonumber\\
&&=
U_6(a_0b_0c_0\qudit{000}+a_0b_0c_1\qudit{112}+a_0b_0c_2\qudit{221}+a_0b_1c_0\qudit{120}+a_0b_1c_1\qudit{202}
\nonumber\\&&\hskip1em
+a_0b_1c_2\qudit{011}+a_0b_2c_0\qudit{210}+a_0b_2c_1\qudit{022}+a_0b_2c_2\qudit{101}+a_1b_0c_0\qudit{201}\nonumber\\&&\hskip1em
+a_1b_0c_1\qudit{010}+a_1b_0c_2\qudit{122}+a_1b_1c_0\qudit{021}+a_1b_1c_1\qudit{100}+a_1b_1c_2\qudit{212}\nonumber\\&&\hskip1em
+a_1b_2c_0\qudit{111}+a_1b_2c_1\qudit{220}+a_1b_2c_2\qudit{002}+a_2b_0c_0\qudit{102}+a_2b_0c_1\qudit{211}\nonumber\\&&\hskip1em
+a_2b_0c_2\qudit{020}+a_2b_1c_0\qudit{222}+a_2b_1c_1\qudit{001}+a_2b_1c_2\qudit{110}+a_2b_2c_0\qudit{012}\nonumber\\&&\hskip1em
+a_2b_2c_1\qudit{121}+a_2b_2c_2\qudit{200}) \nonumber\\
&& =
a_0b_0c_0\qudit{000}+a_0b_0c_1\qudit{012}+a_0b_0c_2\qudit{021}+a_0b_1c_0\qudit{120}+a_0b_1c_1\qudit{102}\nonumber\\&&\hskip1em
+a_0b_1c_2\qudit{111}+a_0b_2c_0\qudit{210}+a_0b_2c_1\qudit{222}+a_0b_2c_2\qudit{201}+a_1b_0c_0\qudit{001}\nonumber\\&&\hskip1em
+a_1b_0c_1\qudit{010}+a_1b_0c_2\qudit{022}+a_1b_1c_0\qudit{121}+a_1b_1c_1\qudit{100}+a_1b_1c_2\qudit{112}\nonumber\\&&\hskip1em
+a_1b_2c_0\qudit{211}+a_1b_2c_1\qudit{220}+a_1b_2c_2\qudit{202}+a_2b_0c_0\qudit{002}+a_2b_0c_1\qudit{011}\nonumber\\&&\hskip1em
+a_2b_0c_2\qudit{020}+a_2b_1c_0\qudit{122}+a_2b_1c_1\qudit{101}+a_2b_1c_2\qudit{110}+a_2b_2c_0\qudit{212}\nonumber\\&&\hskip1em
+a_2b_2c_1\qudit{221}+a_2b_2c_2\qudit{200}.
\end{eqnarray}
Similarly,
\begin{eqnarray}
&&U_7(U_6U_5U_4U_3U_2U_1(a_0b_0c_0\qudit{000}+a_0b_0c_1\qudit{001}+a_0b_0c_2
\qudit{002}+a_0b_1c_0\qudit{010}\nonumber\\&&\hskip1em+a_0b_1c_1\qudit{011}+a_0b_1c_2
\qudit{012}+a_0b_2c_0\qudit{020}+a_0b_2c_1\qudit{021}+a_0b_2c_2\qudit{022}\nonumber\\&&\hskip1em+a_1b_0c_0
\qudit{110}+a_1b_0c_1\qudit{111}+a_1b_0c_2\qudit{112}+a_1b_1c_0\qudit{120}+a_1b_1c_1\qudit{121}\nonumber\\&&\hskip1em
+a_1b_1c_2\qudit{122}+a_1b_2c_0\qudit{100}+a_1b_2c_1\qudit{101}+a_1b_2c_2\qudit{102}+a_2b_0c_0\qudit{220}\nonumber\end{eqnarray}\begin{eqnarray}&&\hskip1em
+a_2b_0c_1\qudit{221}+a_2b_0c_2\qudit{222}+a_2b_1c_0\qudit{200}+a_2b_1c_1\qudit{201}+a_2b_1c_2\qudit{202}\nonumber\\&&\hskip1em
+a_2b_2c_0\qudit{210}+a_2b_2c_1\qudit{211}+a_2b_2c_2\qudit{212}))\nonumber\\
&& =
U_7(a_0b_0c_0\qudit{000}+a_0b_0c_1\qudit{012}+a_0b_0c_2\qudit{021}+a_0b_1c_0\qudit{120}+a_0b_1c_1\qudit{102}
\nonumber\\&&\hskip1em
+a_0b_1c_2\qudit{111}+a_0b_2c_0\qudit{210}+a_0b_2c_1\qudit{222}+a_0b_2c_2\qudit{201}+a_1b_0c_0\qudit{001}\nonumber\\&&\hskip1em
+a_1b_0c_1\qudit{010}+a_1b_0c_2\qudit{022}+a_1b_1c_0\qudit{121}+a_1b_1c_1\qudit{100}+a_1b_1c_2\qudit{112}\nonumber\\&&\hskip1em
+a_1b_2c_0\qudit{211}+a_1b_2c_1\qudit{220}+a_1b_2c_2\qudit{202}+a_2b_0c_0\qudit{002}+a_2b_0c_1\qudit{011}\nonumber\\&&\hskip1em
+a_2b_0c_2\qudit{020}+a_2b_1c_0\qudit{122}+a_2b_1c_1\qudit{101}+a_2b_1c_2\qudit{110}+a_2b_2c_0\qudit{212}\nonumber\\&&\hskip1em
+a_2b_2c_1\qudit{221}+a_2b_2c_2\qudit{200}) \nonumber\\ && =
a_0b_0c_0\qudit{000}+a_0b_0c_1\qudit{012}+a_0b_0c_2\qudit{021}+a_0b_1c_0\qudit{100}+a_0b_1c_1\qudit{112}\nonumber\\&&\hskip1em
+a_0b_1c_2\qudit{121}+a_0b_2c_0\qudit{200}+a_0b_2c_1\qudit{212}+a_0b_2c_2\qudit{221}+a_1b_0c_0\qudit{001}\nonumber\\&&\hskip1em
+a_1b_0c_1\qudit{010}+a_1b_0c_2\qudit{022}+a_1b_1c_0\qudit{101}+a_1b_1c_1\qudit{110}+a_1b_1c_2\qudit{122}\nonumber\\&&\hskip1em
+a_1b_2c_0\qudit{201}+a_1b_2c_1\qudit{210}+a_1b_2c_2\qudit{222}+a_2b_0c_0\qudit{002}+a_2b_0c_1\qudit{011}\nonumber\\&&\hskip1em
+a_2b_0c_2\qudit{020}+a_2b_1c_0\qudit{102}+a_2b_1c_1\qudit{111}+a_2b_1c_2\qudit{120}+a_2b_2c_0\qudit{202}\nonumber\\&&\hskip1em
+a_2b_2c_1\qudit{211}+a_2b_2c_2\qudit{220}
\end{eqnarray}
where the unitary matrix $U_7$ is the matrix given by $U_1$. Our
final gate as outlined in this construction and evident in Fig.
\ref{effswap} is the unitary matrix $U_8$. The unitary matrix
$U_8$ corresponds to the unitary matrix $U_2$, and we have
\begin{figure}$\left(\begin{matrix}
{1\ 0\ 0\ 0\ 0\ 0\ 0\ 0\ 0\ 0\ 0\ 0\ 0\ 0\ 0\ 0\ 0\ 0\ 0\ 0\ 0\ 0\
0\ 0\ 0\ 0\ 0}\cr {0\ 0\ 0\ 0\ 0\ 0\ 0\ 0\ 0\ 1\ 0\ 0\ 0\ 0\ 0\ 0\
0\ 0\ 0\ 0\ 0\ 0\ 0\ 0\ 0\ 0\ 0}\cr {0\ 0\ 0\ 0\ 0\ 0\ 0\ 0\ 0\ 0\
0\ 0\ 0\ 0\ 0\ 0\ 0\ 0\ 1\ 0\ 0\ 0\ 0\ 0\ 0\ 0\ 0}\cr {0\ 1\ 0\ 0\
0\ 0\ 0\ 0\ 0\ 0\ 0\ 0\ 0\ 0\ 0\ 0\ 0\ 0\ 0\ 0\ 0\ 0\ 0\ 0\ 0\ 0\
0}\cr {0\ 0\ 0\ 0\ 0\ 0\ 0\ 0\ 0\ 0\ 1\ 0\ 0\ 0\ 0\ 0\ 0\ 0\ 0\ 0\
0\ 0\ 0\ 0\ 0\ 0\ 0}\cr {0\ 0\ 0\ 0\ 0\ 0\ 0\ 0\ 0\ 0\ 0\ 0\ 0\ 0\
0\ 0\ 0\ 0\ 0\ 1\ 0\ 0\ 0\ 0\ 0\ 0\ 0}\cr {0\ 0\ 1\ 0\ 0\ 0\ 0\ 0\
0\ 0\ 0\ 0\ 0\ 0\ 0\ 0\ 0\ 0\ 0\ 0\ 0\ 0\ 0\ 0\ 0\ 0\ 0}\cr {0\ 0\
0\ 0\ 0\ 0\ 0\ 0\ 0\ 0\ 0\ 1\ 0\ 0\ 0\ 0\ 0\ 0\ 0\ 0\ 0\ 0\ 0\ 0\
0\ 0\ 0}\cr {0\ 0\ 0\ 0\ 0\ 0\ 0\ 0\ 0\ 0\ 0\ 0\ 0\ 0\ 0\ 0\ 0\ 0\
0\ 0\ 1\ 0\ 0\ 0\ 0\ 0\ 0}\cr {0\ 0\ 0\ 1\ 0\ 0\ 0\ 0\ 0\ 0\ 0\ 0\
0\ 0\ 0\ 0\ 0\ 0\ 0\ 0\ 0\ 0\ 0\ 0\ 0\ 0\ 0}\cr {0\ 0\ 0\ 0\ 0\ 0\
0\ 0\ 0\ 0\ 0\ 0\ 1\ 0\ 0\ 0\ 0\ 0\ 0\ 0\ 0\ 0\ 0\ 0\ 0\ 0\ 0}\cr
{0\ 0\ 0\ 0\ 0\ 0\ 0\ 0\ 0\ 0\ 0\ 0\ 0\ 0\ 0\ 0\ 0\ 0\ 0\ 0\ 0\ 1\
0\ 0\ 0\ 0\ 0}\cr {0\ 0\ 0\ 0\ 1\ 0\ 0\ 0\ 0\ 0\ 0\ 0\ 0\ 0\ 0\ 0\
0\ 0\ 0\ 0\ 0\ 0\ 0\ 0\ 0\ 0\ 0}\cr {0\ 0\ 0\ 0\ 0\ 0\ 0\ 0\ 0\ 0\
0\ 0\ 0\ 1\ 0\ 0\ 0\ 0\ 0\ 0\ 0\ 0\ 0\ 0\ 0\ 0\ 0}\cr {0\ 0\ 0\ 0\
0\ 0\ 0\ 0\ 0\ 0\ 0\ 0\ 0\ 0\ 0\ 0\ 0\ 0\ 0\ 0\ 0\ 0\ 1\ 0\ 0\ 0\
0}\cr {0\ 0\ 0\ 0\ 0\ 1\ 0\ 0\ 0\ 0\ 0\ 0\ 0\ 0\ 0\ 0\ 0\ 0\ 0\ 0\
0\ 0\ 0\ 0\ 0\ 0\ 0}\cr {0\ 0\ 0\ 0\ 0\ 0\ 0\ 0\ 0\ 0\ 0\ 0\ 0\ 0\
1\ 0\ 0\ 0\ 0\ 0\ 0\ 0\ 0\ 0\ 0\ 0\ 0}\cr {0\ 0\ 0\ 0\ 0\ 0\ 0\ 0\
0\ 0\ 0\ 0\ 0\ 0\ 0\ 0\ 0\ 0\ 0\ 0\ 0\ 0\ 0\ 1\ 0\ 0\ 0}\cr {0\ 0\
0\ 0\ 0\ 0\ 1\ 0\ 0\ 0\ 0\ 0\ 0\ 0\ 0\ 0\ 0\ 0\ 0\ 0\ 0\ 0\ 0\ 0\
0\ 0\ 0}\cr {0\ 0\ 0\ 0\ 0\ 0\ 0\ 0\ 0\ 0\ 0\ 0\ 0\ 0\ 0\ 0\ 1\ 0\
0\ 0\ 0\ 0\ 0\ 0\ 0\ 0\ 0}\cr {0\ 0\ 0\ 0\ 0\ 0\ 0\ 0\ 0\ 0\ 0\ 0\
0\ 0\ 0\ 0\ 0\ 0\ 0\ 0\ 0\ 0\ 0\ 0\ 1\ 0\ 0}\cr {0\ 0\ 0\ 0\ 0\ 0\
0\ 1\ 0\ 0\ 0\ 0\ 0\ 0\ 0\ 0\ 0\ 0\ 0\ 0\ 0\ 0\ 0\ 0\ 0\ 0\ 0}\cr
{0\ 0\ 0\ 0\ 0\ 0\ 0\ 0\ 0\ 0\ 0\ 0\ 0\ 0\ 0\ 0\ 0\ 1\ 0\ 0\ 0\ 0\
0\ 0\ 0\ 0\ 0}\cr {0\ 0\ 0\ 0\ 0\ 0\ 0\ 0\ 0\ 0\ 0\ 0\ 0\ 0\ 0\ 0\
0\ 0\ 0\ 0\ 0\ 0\ 0\ 0\ 0\ 1\ 0}\cr {0\ 0\ 0\ 0\ 0\ 0\ 0\ 0\ 1\ 0\
0\ 0\ 0\ 0\ 0\ 0\ 0\ 0\ 0\ 0\ 0\ 0\ 0\ 0\ 0\ 0\ 0}\cr {0\ 0\ 0\ 0\
0\ 0\ 0\ 0\ 0\ 0\ 0\ 0\ 0\ 0\ 0\ 0\ 0\ 0\ 1\ 0\ 0\ 0\ 0\ 0\ 0\ 0\
0}\cr {0\ 0\ 0\ 0\ 0\ 0\ 0\ 0\ 0\ 0\ 0\ 0\ 0\ 0\ 0\ 0\ 0\ 0\ 0\ 0\
0\ 0\ 0\ 0\ 0\ 0\ 1}\cr
\end{matrix}\right)$\caption{The qutrit SWAP matrix}
\end{figure}
\begin{eqnarray}
&&U_8(U_7U_6U_5U_4U_3U_2U_1(a_0b_0c_0\qudit{000}+a_0b_0c_1\qudit{001}+a_0b_0c_2\qudit{002}
+a_0b_1c_0\qudit{010}\nonumber\\&&\hskip1em+a_0b_1c_1\qudit{011}+a_0b_1c_2\qudit{012}+a_0b_2c_0\qudit{020}
+a_0b_2c_1\qudit{021}+a_0b_2c_2\qudit{022}\nonumber\\&&\hskip1em+a_1b_0c_0\qudit{110}+a_1b_0c_1\qudit{111}
+a_1b_0c_2\qudit{112}+a_1b_1c_0\qudit{120}+a_1b_1c_1\qudit{121}\nonumber\\&&\hskip1em+a_1b_1c_2\qudit{122}
+a_1b_2c_0\qudit{100}+a_1b_2c_1\qudit{101}+a_1b_2c_2\qudit{102}+a_2b_0c_0\qudit{220}\nonumber\\&&
\hskip1em+a_2b_0c_1\qudit{221}+a_2b_0c_2\qudit{222}+a_2b_1c_0\qudit{200}+a_2b_1c_1\qudit{201}+a_2b_1c_2\qudit{202}
\nonumber\\&&\hskip1em+a_2b_2c_0\qudit{210}+a_2b_2c_1\qudit{211}+a_2b_2c_2\qudit{212}))\nonumber\\
&& =
U_8(a_0b_0c_0\qudit{000}+a_0b_0c_1\qudit{012}+a_0b_0c_2\qudit{021}+a_0b_1c_0\qudit{100}+a_0b_1c_1
\qudit{112}\nonumber\\&&\hskip1em
+a_0b_1c_2\qudit{121}+a_0b_2c_0\qudit{200}+a_0b_2c_1\qudit{212}+a_0b_2c_2
\qudit{221}+a_1b_0c_0\qudit{001}\nonumber\\&&\hskip1em+a_1b_0c_1\qudit{010}
+a_1b_0c_2\qudit{022}+a_1b_1c_0\qudit{101}+a_1b_1c_1\qudit{110}+a_1b_1c_2\qudit{122}\nonumber\\&&
\hskip1em+a_1b_2c_0\qudit{201}+a_1b_2c_1\qudit{210}+a_1b_2c_2\qudit{222}+a_2b_0c_0\qudit{002}+a_2b_0c_1
\qudit{011}\nonumber\\&&\hskip1em+a_2b_0c_2\qudit{020}+a_2b_1c_0\qudit{102}+a_2b_1c_1\qudit{111}+a_2b_1c_2
\qudit{120}+a_2b_2c_0\qudit{202}\nonumber\\&&\hskip1em+a_2b_2c_1\qudit{211}+a_2b_2c_2\qudit{220})
\nonumber\\ && =
a_0b_0c_0\qudit{000}+a_0b_0c_1\qudit{010}+a_0b_0c_2\qudit{020}+a_0b_1c_0\qudit{100}+a_0b_1c_1\qudit{110}\nonumber\\&&\hskip1em
+a_0b_1c_2\qudit{120}+a_0b_2c_0\qudit{200}+a_0b_2c_1\qudit{210}+a_0b_2c_2\qudit{220}+a_1b_0c_0\qudit{001}\nonumber\\&&\hskip1em
+a_1b_0c_1\qudit{011}+a_1b_0c_2\qudit{021}+a_1b_1c_0\qudit{101}+a_1b_1c_1\qudit{111}+a_1b_1c_2\qudit{121}\nonumber\\&&\hskip1em
+a_1b_2c_0\qudit{201}+a_1b_2c_1\qudit{211}+a_1b_2c_2\qudit{221}+a_2b_0c_0\qudit{002}+a_2b_0c_1\qudit{012}\nonumber\\&&\hskip1em
+a_2b_0c_2\qudit{022}+a_2b_1c_0\qudit{102}+a_2b_1c_1\qudit{112}+a_2b_1c_2\qudit{122}+a_2b_2c_0\qudit{202}\nonumber\\&&\hskip1em
+a_2b_2c_1\qudit{212}+a_2b_2c_2\qudit{222}.
\end{eqnarray}

\noindent Now, the state of the system after application of the
final unitary matrix $U_8$ may be written
\begin{eqnarray}
&&a_0b_0c_0\qudit{000}+a_0b_0c_1\qudit{010}+a_0b_0c_2\qudit{020}+a_0b_1c_0\qudit{100}+a_0b_1c_1\qudit{110}\nonumber\\
&&+a_0b_1c_2\qudit{120}+a_0b_2c_0\qudit{200}+a_0b_2c_1\qudit{210}+a_0b_2c_2\qudit{220}+a_1b_0c_0\qudit{001}\nonumber\end{eqnarray}\begin{eqnarray}
&&\hskip1em+a_1b_0c_1\qudit{011}+a_1b_0c_2\qudit{021}+a_1b_1c_0\qudit{101}+a_1b_1c_1\qudit{111}+a_1b_1c_2\qudit{121}\nonumber\\
&&\hskip1em+a_1b_2c_0\qudit{201}+a_1b_2c_1\qudit{211}+a_1b_2c_2\qudit{221}+a_2b_0c_0\qudit{002}+a_2b_0c_1\qudit{012}\nonumber\\
&&\hskip1em+a_2b_0c_2\qudit{022}+a_2b_1c_0\qudit{102}+a_2b_1c_1\qudit{112}+a_2b_1c_2\qudit{122}+a_2b_2c_0\qudit{202}\nonumber\\
&&\hskip1em+a_2b_2c_1\qudit{212}+a_2b_2c_2\qudit{222}\nonumber\\
&& =
b_0c_0a_0\qudit{000}+b_0c_0a_1\qudit{001}+b_0c_0a_2\qudit{002}+b_0c_1a_0\qudit{010}+b_0c_1a_1\qudit{011}\nonumber\\
&&\hskip1em+b_0c_1a_2\qudit{012}+
b_0c_2a_0\qudit{020}+b_0c_2a_1\qudit{021}+b_0c_2a_2\qudit{022}+b_1c_0a_0\qudit{100}\nonumber
\\
&&\hskip1em+b_1c_0a_1\qudit{101}+b_1c_0a_2\qudit{102}+ b_1c_1a_0\qudit{110}+b_1c_1a_1\qudit{111}+b_1c_1a_2\qudit{112}\nonumber\\
&&\hskip1em+b_1c_2a_0\qudit{121}+b_1c_2a_1\qudit{121}+b_1c_2a_2\qudit{122}+
b_2c_0a_0\qudit{200}+b_2c_0a_1\qudit{201}\nonumber\\
&&\hskip1em+b_2c_0a_2\qudit{202}+b_2c_1a_0\qudit{210}+b_2c_1a_1\qudit{211}+b_2c_1a_2\qudit{212}+b_2c_2a_0\qudit{220}\nonumber\\
&&\hskip1em+b_2c_2a_1\qudit{221}+b_2c_2a_2\qudit{222}.
\label{last1}\end{eqnarray} In particular, we note that state
\ref{last1}  has the form \begin{eqnarray}
&&(b_0\qudit{0}+b_1\qudit{1}+b_2\qudit{2})\otimes (c_0\qudit{0}+c_1\qudit{1}+c_2\qudit{2})\otimes (a_0\qudit{0}+a_1\qudit{1}+a_2\qudit{2})\nonumber\\
&&\ \ \  = \qudit{b}\otimes\qudit{c}\otimes\qudit{a},
\end{eqnarray}

\noindent thereby illustrating our binomial summation construction
to be a quantum {\small{SWAP}} gate.


\end{document}